\documentclass[showpacs,preprintnumbers,amsmath,amssymb]{revtex4}
\usepackage{graphicx}
\usepackage{dcolumn}
\usepackage{bm}

\newcommand{\be}{\begin{equation}}
\newcommand{\ee}{\end{equation}}
\newcommand{\ba}{\begin{eqnarray}}
\newcommand{\ea}{\end{eqnarray}}
\newcommand{\bas}{\begin{eqnarray*}}
\newcommand{\eas}{\end{eqnarray*}}
\newcommand{\nn}{\nonumber \\}

\newcommand{\edth}{\,\eth\,}
\newcommand{\edthbar}{\,\overline{\eth}\,}

\newcommand{\nhatb}{\mbox{\boldmath $\hat{\n}$}}
\newcommand{\thehatb}{\mbox{\boldmath $\hat{\theta}$}}

\newcommand{\varphihatb}{\mbox{\boldmath $\hat{\varphi}$}}

\newcommand{\eI}{\mbox{\boldmath $\hat{e}_1$}}
\newcommand{\eII}{\mbox{\boldmath $\hat{e}_2$}}

\newcommand{\etheta}{\mbox{\boldmath $\hat{e}_{\theta}$}}
\newcommand{\ephi}{\mbox{\boldmath $\hat{e}_{\varphi}$}}

\newcommand{\ex}{\mbox{\boldmath $\hat{e}_{x}$}}
\newcommand{\ey}{\mbox{\boldmath $\hat{e}_{y}$}}

\newcommand{\m}{\,\mbox{\boldmath $m$}}
\newcommand{\mbar}{\,\mbox{\boldmath $\overline{m}$}}

\newcommand{\sfunc}{\,_sf}
\newcommand{\smfunc}{\,_{-s}f}
\newcommand{\spmfunc}{\,_{\pm s}f}

\newcommand{\rr}{\mbox{\boldmath $r$}}
\newcommand{\n}{\mbox{\boldmath $n$}}
\newcommand{\vb}{\mbox{\boldmath $v$}}
\newcommand{\x}{\mbox{\boldmath $x$}}
\newcommand{\kk}{\mbox{\boldmath $k$}}

\newcommand{\vth}{\vec{\theta}}
\newcommand{\vell}{\vec{\ell}}

\newcommand{\de}{\partial}
\newcommand{\nablat}{\nabla_{\theta}}
\newcommand{\nablap}{\nabla_{\varphi}}

\newcommand{\rgl}{\rangle}
\newcommand{\lgl}{\langle}

\newcommand{\lm}{{\ell m}}
\newcommand{\lmd}{{\ell' m'}}

\newcommand{\lld}{{\ell \ell'}}
\newcommand{\mmd}{{m m'}}
\newcommand{\Ylm}{Y_\lm}
\newcommand{\sYlm}{\,_sY_\lm}

\newcommand{\tYlm}{\,_2\!Y_\lm}
\newcommand{\mtYlm}{\,_{-2}\!Y_\lm}
\newcommand{\pmtYlm}{\,_{\pm 2}\!Y_\lm}

\newcommand{\Zklm}{\,Z_{k\lm}}
\newcommand{\sZklm}{\,_sZ_{k\lm}}

\newcommand{\ZklmpII}{\,_2Z_{k\lm}}
\newcommand{\ZklmmII}{\,_{-2}Z_{k\lm}}
\newcommand{\ZklmpmII}{\,_{\pm2}Z_{k\lm}}

\newcommand{\AaA}{Astron. Astrophys. }
\newcommand{\AJ}{Astrophys. J. }
\newcommand{\AJS}{Astrophys. J. Supp. }

\newcommand{\AJL}{Astrophys. J. Lett. }
\newcommand{\ARAA}{Ann. Rev. Astron. Astrophys. }
\newcommand{\MNRAS}{Mon. Not. Roy. Astron. Soc. }

\newcommand{\PR}{Phys. Rep. }

\newcommand{\PRD}{Phys. Rev. D }

\newcommand{\PRL}{Phys. Rev. Lett. }
\newcommand{\JMP}{J. Math. Phys. }
\newcommand{\RMP}{Rev. Mod. Phys. }

\begin{document}

\preprint{APS/000-000}

\title{Weak lensing analysis in three dimensions}

\author{P. G.
Castro\footnote{pgc@roe.ac.uk;$\,^\dagger$afh@roe.ac.uk;$\,^\ddag$tdk@roe.ac.uk},
A. F. Heavens$\,^\dagger$ and T. D. Kitching$\,^\ddag$}
\affiliation{%
Institute for Astronomy, University of Edinburgh,
Royal Observatory, Blackford Hill, Edinburgh EH9 3HJ, U.K.
}

\date{\today}

\begin{abstract}
We present a comprehensive full-sky 3-dimensional analysis of the
weak-lensing fields and their corresponding power spectra. Using
the formalism of spin-weight spherical harmonics and spherical
Bessel functions, we relate the two-point statistics of the
harmonic expansion coefficients of the weak lensing shear and
convergence to the power spectrum of the matter density
perturbations, and derive small-angle limits. Such a study is
relevant in view of the next generation of large-scale
weak lensing surveys which will provide distance information about
the sources through photometric redshifts.  This opens up the
possibility of accurate cosmological parameter estimation via weak
lensing, with an emphasis on the equation of state of dark energy.
\end{abstract}

\maketitle

\section{Introduction}
\label{sec:intro}

Like the Cosmic Microwave Background (CMB) a decade ago, the study
of gravitational weak lensing on a cosmic scale is now entering a
promised golden age. As with the CMB, it is well understood
theoretically~\cite{art:BartSchneider99,art:WaerbekeMellier03}, it
has now been detected (see
e.g.~\cite{art:BaconEtAl00,art:WaerbekeEtAl01,art:HoekstraEtAl02,art:HamanaEtAl02,art:BrownEtAl03,art:JarvisEtAl03})
and ambitious experimental projects are under way. 
The uniqueness and thus the appeal of weak
lensing lies in its clean and direct sensitivity to the total mass
distribution of the Universe. Indeed, it arises from the
deflection of light by the gravitational potential of the matter
along the photon path, regardless of the precise nature or state
of the intervening matter. With weak lensing one can hope to
measure for the first time the 3D matter power spectrum over a
wide range of scales, in principle independently of any model for
evolution. It thus provides complementary information to that
rendered by the CMB alone or other lower redshift cosmological
probes. The CMB does provide the cleanest possible window into the
physics of the early Universe and its
parameters~\cite{art:HuDodelson02,art:SpergetEtAl03}, but lacks
strong sensitivity to the Universe's subsequent evolution.
Large-scale galaxy surveys are hindered by the presence of
non-linearities which affect the luminous baryonic matter on small
scales and have to rely on assumptions about (or deductions of)
the relationship between luminous matter and mass distribution on
large
scales~\cite{art:VerdeEtAl02,art:ColeEtAl05,art:EisensteinEtAl05}.
Supernova studies are afflicted by complicated
systematics~\cite{art:AlderingEtAl04}. Hence, weak lensing can
considerably help in the determination of poorly constrained
parameters linked to the matter evolution and the
moderate-redshift evolution of the redshift-distance relation. The
most significant parameter is the equation of state of dark
energy, $w\equiv p / \rho c^2$, and its time evolution. As is now
generally accepted, there is strong evidence from a variety of
cosmological probes like supernovae~\cite{art:PerlmutterEtAl99}
and the CMB~\cite{art:SpergetEtAl03} that a dark energy component
makes up about $70$ \% of the total energy density budget of our
Universe but its precise nature and state remain a mystery. Answers
are unlikely to come from particle physics and our hopes lie in
cosmology.

A variety of deep, large sky weak lensing surveys are either
ongoing, scheduled or planned in the foreseeable future (Deep Lens
Survey~\cite{web:DLS}, NOAO Deep survey~\cite{web:NOAO}, The CFHT
Legacy survey~\cite{web:CFHT}, darkCAM on VISTA, The Panoramic
Survey Telescope and Rapid response System
(Pan-STARRS)~\cite{web:panstarr}, the Large Synoptic Survey
Telescope~\cite{web:LSST} and SNAP~\cite{web:SNAP}).  If one can
rigorously account for the various systematics (such as the
seeing, calibration, anisotropic point spread functions, redshift
uncertainties etc.) likely to permeate the data, such surveys can
offer if not definite answers at least fundamental clues on the
nature of the dark energy, which lensing surveys are beginning to
constrain $w$~\cite{art:Jarvisetal05}. In combination with other
data sets, weak lensing surveys data can further overcome
intrinsic degeneracies characteristic of each of the cosmological
probes and strengthen the constraints on our cosmological model.
In particular, predictive studies have been done which explore the
possible improvements for parameter estimation of considering the
three dimensionality of weak lensing, where photometric redshifts
are used to provide estimates in three dimensions of the weak
lensing shear field. As shown by Ishak~\cite{art:Ishak05}, cosmic
complementarity and 3D lensing tomography bring uncertainties on
the equation of state to the level of a few percent. The use of
full distance information on weak lensing surveys, enabling a full
3D analysis rather than tomography, can further reduce statistical
errors on cosmological parameters~\cite{art:Heavens03}.

Besides reducing statistical errors, there are other reasons for
wishing to have photometric redshifts: weak lensing studies have to
deal with systematic errors which could dominate the error budget
as surveys improve (see~\cite{art:Heavens03} and references
therein). The most important of these is the redshift distribution
of the sources but correlated source galaxy intrinsic
ellipticities (to date still not well-known) and the effect of
source clustering are likely to become a considerable nuisance in
the future. A reasonably accurate knowledge of the redshift of the
sources can help in understanding, quantifying and ultimately
removing such systematics. Furthermore, since acquisition of
photometric redshifts requires several images of each galaxy in
different bands, there is scope for better shape measurement,
independent lensing studies, and so on.

In the past, most of weak lensing analysis were essentially
limited to 2D approaches due to the lack of, or imprecise,
photometric redshift information about individual galaxies.
Moreover, such studies were usually applied to small patches of
the sky necessarily involving a flat-sky approximation (see
e.g.~\cite{art:KamionEtAl98,art:CrittendenEtAl00} 
and~\cite{art:WaerbekeEtAl01,art:BrownEtAl03} for an application to
data). A useful exception to this early trend is the theoretical
work by Stebbins~\cite{art:Stebbins96} who explored the 2D curved
sky by means of the tensor spherical harmonics formalism.
But over the last few years or so, with the arrival of surveys
with more accurate photometric redshift information and the
prospect of them reaching wider areas of the sky, interest has
steadily grown in exploring the inherently full-sky 3D information
contained in the lensing observables (see
e.g.~\cite{art:CouchmanEtAl99,art:BarberEtAl00,art:BaconEtAl04}).
The way of dealing with any source redshift information has thus
evolved over the years from basically a black and white picture to
a `fabulous Technicolour' approach. Originally it was used as a
way of determining the redshift distribution of the sources only.
It then evolved to a so-called tomographic analysis, where the
sources were divided up in slices at different redshifts and a 2D
analysis was then performed on each one of the slices. This can be
useful when trying to detect clusters of
galaxies~\cite{art:WittmanEtAl01,art:WittmanEtAl03,art:TaylorEtAl03,art:BaconEtAl04}.
As shown by Hu~\cite{art:Hu99} at a statistical level such method
also presents certain advantages depending on the parameters one
is interested in retrieving. For instance, for the amplitude of
the matter power spectrum there can be significant gains when a
source population is split into two, but little is gained by
further finer subdivisions. For $w$ the gains are much larger as
the 2D analysis constraints are weak. It is worth noting here that
other 2D routes have been explored in pursuit of cosmologically
sensitive weak-lensing estimators such as the study of
higher-order statistics (for bispectrum detections
see~\cite{art:BernardeauEtAl02,art:PenEtAl03}) but if one wants to
place strong constraints on $w$ a 3D study is compulsory. Several
theoretical
studies~\cite{art:Taylor01,art:HuKeeton03,art:BaconTaylor02,art:Hu02}
investigated how the lensing equations can be inverted to extract
the 3D gravitational potential directly, thus offering the
possibility of reconstruction of the 3D mass density field from
weak-lensing data. Only recently in Heavens~\cite{art:Heavens03},
was a truly 3D spectral statistical analysis on the full-sky
developed, in which the individual source redshifts were taken
into consideration {\it per se}. In there, the attention was
mainly focused upon understanding the amount of extra information
on the cosmological parameters that can be gleaned from a 3D shear
map, rather than on establishing a framework for 3D weak lensing
studies. Motivated by the previous theoretical considerations and
the upcoming experimental prospects, it is timely to develop in
detail a description of the weak lensing observables on the full
three dimensional sky.

A full-sky 3D study of the weak lensing observables needs to
combine at its root both the 3D weak lensing characteristics and
the full-sky formalism. The weak lensing effect can be observed as
a local modification of the surface number density of background
galaxies (the magnification) and a distortion of their shape (the
shear). As has been pointed out repeatedly before, the weak shear
components present striking similarities with the Stokes
parameters $Q$ and $U$ describing the linear polarisation of the
CMB light, the major difference being its additional radial
dependence. Like them, they are not invariant under a rotation of
the reference frame used to describe them but rather transform as
rank-2 tensors. One can then legitimately extend to the third
radial dimension past work done for the full-sky polarisation of
the
CMB~\cite{art:ZaldSeljak97,art:Hu00,art:LewisEtAl01,art:BunnEtAl02,art:OkamotoHu03}
and adapt it to the weak lensing analysis. In this work, we
develop in more detail Heavens' original 3D full-sky
analysis~\cite{art:Heavens03} weaving into our approach both the
CMB polarisation formalism and past weak lensing theoretical 2D
studies~\cite{art:Stebbins96,art:KamionEtAl98,art:CrittendenEtAl00}.
We use a spectral decomposition as it allows one to restrict the
analysis to, for example, the linear or mildly non-linear regime
and rely on the spin-weighted formalism, originally developed by
Newman and Penrose in the 1960s~\cite{art:NewmanPenrose66}.

This paper is organised as follows. In
section~\ref{sec:wl_theory}, we review the main results of the
theory of gravitational weak lensing which we generalise to 3
dimensions so that the relevant theoretical expressions depend
explicitly on the radial distance from us. In
subsection~\ref{subsec:wl_tens} we start by presenting the weak
lensing theoretical expressions in the familiar tensorial
formalism, both in the full-sky and in the flat-sky approximation.
In subsection~\ref{subsec:wl_edth} we introduce an alternative
differential formalism we name the {\it edth} formalism, which
encompasses the full and the flat-sky geometries. Such a
mathematical apparatus reveals itself to be extremely convenient
when working on the full-sky and has been extensively used in
studies of the CMB polarisation but which we develop here for 3D
weak-lensing fields. In section~\ref{sec:wl_full}, we
spectrally decompose in the 3D full-sky the (spin-weight 2) weak
lensing shear and (scalar) convergence fields in terms of
spin-weighted spherical harmonics and Bessel functions. We derive
their expansion coefficients in function of the gravitational
potential coefficients by making full use of the {\it edth}
formalism. In
subsections~\ref{subsec:wl_power_spectra_full} and
\ref{subsec:wl_power_spectra_flat}, we calculate the weak lensing
3D power spectra in the full-sky and derive small-angle limits for
completeness. Finally in section~\ref{sec:3D_corr_function} we present
results for the 2-point correlation function of the 3D shear field.
In the Appendixes, we review in more detail the
notation, the mathematics and the conventions chosen of the
spin-weight $s$ functions and of the associated geometrical spin
raising and lowering ($\edth$ and $\edthbar$) operators defined
over any two-dimensional Riemannian manifold.


\section{Weak lensing theory in 3D}
\label{sec:wl_theory}

\subsection{The lensing potential}
\label{subsec:grav_pot}

In the linear regime, many scalar fields on the sky that are
associated with large-scale structures can be interpreted as
line-of-sight integrations of functions of the gravitational
potential $\Phi$ with a given weight. A few examples of such
scalar fields are the Integrated Sachs-Wolfe
effect~\cite{art:SachsWolfe67} or the Ostriker-Vishniac
effect~\cite{art:Vishniac87} imprinting themselves on the CMB and
the gravitational weak lensing by cosmological structure (for a
comprehensive review see~\cite{art:BartSchneider99}). For the
gravitational weak lensing case, one can associate the so-called
weak lensing potential $\phi$ for a given source at a 3D position
in comoving space $\rr \equiv (r,\theta,\varphi)$ to the peculiar
gravitational potential $\Phi$ defined along the line-of-sight via
(see~\cite{art:BartSchneider99}) \ba \phi (\rr) & = &  \phi
(r,\theta, \varphi) = \frac{2}{c^2}
                                     \int_{0}^{r}
                                     dr'\frac{f_K(r-r')}{f_K(r)f_K(r')}
                                     \Phi(r',\theta,\varphi)
\label{eq:pot} \ea where the Born approximation was assumed (i.e.
the path of the integration, corresponding to the path of the
photons emitted by the source, is assumed to be unperturbed by the
lens). Although usually the lensing potential $\phi$ is regarded
as a 2D radial projection of the 3D gravitational potential, it is
in reality a 3D quantity. It is customary to average over the
redshift distribution of the source galaxies, but this is not
necessary if one has distance information about the individual
sources. Here and in the remainder of the manuscript, bold letters
denote 3-dimensional vectors, $c$ is the speed of light, $r \equiv
r(t)$ is the comoving distance of the source at instant $t$ from
the observer at the origin ($r=0$) and $f_K(r)d\psi$ is the
comoving transverse dimensionless separation for points separated
by $d\psi$. We have $f_K(r)={\sin r}, r$ and ${\sinh r}$ for a
closed ($k=1$), flat ($k=0$) and open ($k=-1$) universes
respectively. The gravitational potential $\Phi$ is related to the
underlying overdensity field $\delta(\rr)\equiv \delta \rho
(\rr)/{\overline \rho}$ by the Poisson equation 
\be 
\nabla_r^2
\Phi(\rr) = \frac{3\Omega_m\,H_0^2}{2a(t)}\delta(\rr)
\label{eq:poisson} 
\ee 
where the 3D gradient $\nabla_r$ is defined
relative to comoving coordinates, $\Omega_m$ is the present-day
total matter density, $H_0$ is the Hubble constant today in units
of km/s/Mpc and $a(t)=1/(1+z)$ is the scale factor. We use the
comoving gauge or total-matter gauge (see e.g.~\cite{book:LiddleLyth00,art:MaBert95}).

If one wants to make a spectral expansion, such as a Fourier
transform, then immediately one has a subtlety to consider.  The
lensing potential field, indirectly observed through its effect on
the shapes of lensed galaxies, is not homogeneous, as it is viewed
on the past light cone of the galaxies, and is a function of the
gravitational potential which evolves with cosmic epoch.  When
referring to the transform of a field, such as the gravitational
potential, at a lensed galaxy at distance $r$, we will mean the
transform of the {\em homogeneous} field existing everywhere at
the cosmic epoch corresponding to the time the observed light left
the distant galaxy.  Thus the coefficients of the expansion depend
on the lookback time of the observation, and hence, rather
paradoxically, on the distance $r$ itself.

There is a natural choice of basis functions to use for an
expansion. Spherical coordinates are natural for various reasons,
partly because the lensing potential is a radial integral, partly
because partial sky coverage is more easily dealt with, as
demonstrated in large-scale structure studies by
\cite{art:HeavensTaylor95}. In addition, errors in distances (from
using photometric redshifts as distance indicators, for instance)
are radial errors. The first of these means the coefficients of
the expansion of the gravitational potential and of the lensing
potential are straightforwardly related. The choice of basis
functions is motivated by Poisson's equation (\ref{eq:poisson}).
It makes sense to use the eigenfunctions of the Laplacian
operator, since then the coefficients of the gravitational
potential and of the density field are closely related
(essentially by a factor $-k^{-2}$).  In Cartesian coordinates,
the eigenfunctions are the familiar exponential functions of
Fourier analysis. If however the Laplacian is written in spherical
coordinates ($r,\theta,\varphi$), then in flat space the
eigenfunctions become products of spherical harmonics and
spherical Bessel functions: $Y_{\ell
m}(\theta,\varphi)j_\ell(kr)$, with eigenvalue $-k^2$. So for
scalar fields $f(\rr)$ in a flat background geometry, the natural
3D expansion is
\begin{equation}
f_{\ell m}(k) \equiv \sqrt{2\over \pi} \int d^3 r f(\rr)
                   k \,j_\ell(kr) Y_{\ell m}^*(\theta,\varphi),
\label{eq:general_3D_expansion_lm}
\end{equation}
where the numerical factor and the presence of $k$ are chosen for
convenience. The inverse transform is
\begin{equation}
f(\rr) = \sqrt{2\over \pi} \int k \,dk \sum_{\ell=0}^\infty
         \sum_{m=-\ell}^\ell f_{\ell m}(k) j_\ell(k r) Y_{\ell m}(\theta,
\varphi). 
\label{eq:general_3D_expansion}
\end{equation}
This is readily obtained from the orthonormality of the spherical
harmonics and the orthogonality of the spherical Bessel 
functions (see Eqs.~(\ref{eq:sph_harm_orth_comp_1}) and (\ref{eq:orth_bessel})).

Using this 3-dimensional expansion,
one can relate the transform of the lensing potential to the
gravitational potential field in a flat geometry~\cite{art:Heavens03} by
\be
\phi_{\lm}(k) = \frac{4 k}{\pi c^2}
                  \int_{0}^{\infty} dk'\, k' \int_{0}^{\infty}dr\,r\, j_\ell(kr)
                  \int_{0}^{r} dr'
                 \left[ \frac{r-r'}{r'}\right]
                  j_{\ell}(k'r') \Phi_{\lm}(k';r').
\label{eq:transf_lens_grav} \ee It is worth remarking that our
expression differs from Eqs.~(6) and (7) found in
Heavens~\cite{art:Heavens03} due to a different choice of the 3D
expansion conventions. One can further relate $\Phi_{\lm}(k;r)$ to
the expansion of the matter overdensity $\delta_{\lm}(k;r)$ by
using the Poisson equation:
\begin{equation}
 \Phi_{\lm}(k; r) = -
\frac{3\Omega_m H_0^2}{2k^2 a(r)}\,\delta_{\lm}(k; r),
\label{eq:poisson_bis}
\end{equation}
where here we put the time dependence (see discussion above) in
the coefficients of the gravitational potential and matter
overdensity fields by explicitly writing the coefficients as
functions of the distance $r$ in addition to $\ell$, $m$ and $k$. We point out
that the expansion coefficients of the lensing potential $\phi$ do
not have the time dependence explicitly shown because $\phi$ is
not by definition a homogeneous and isotropic field in 3D space.

These equations establish the relations between the scalar fields
$\phi$, $\Phi$ and $\delta$.  Since the statistical properties of
$\delta$ are known for a given cosmological model, this opens up
the possibility of using weak lensing in 3D (via $\phi$) for
estimation of cosmological parameters.  This would normally be
done using the weak lensing shear field, $\gamma(\rr)$, which is
not a scalar field, but rather a spin-weight 2 field.  We will
consider the statistics of this observable field in Section III.

\subsection{Weak lensing and the gravitational potential in the
tensorial formalism}
\label{subsec:wl_tens}

In this subsection we review the main results of the theory of
gravitational weak lensing in the tensorial formalism on the 3
dimensional spherical sky. In addition, we introduce the relevant
formulae in the 3 dimensional flat-sky approximation for
completeness.

\subsubsection{Weak lensing on the full-sky}

The 2 dimensional distortion of images of distant sources, located at a
certain 3D comoving position in space $\rr$, caused by the
weak gravitational lensing by intervening structures
is given by~\cite{art:BartSchneider99,art:BernardeauEtAl97,art:Hu00,art:Hu01}
\ba
[\nabla_i \nabla_j - \frac{1}{2}g_{ij}\nabla^2] \phi (\rr)& = &
            [\gamma_1(\rr) \sigma_3 + \gamma_2(\rr) \sigma_1]_{ij}
\label{eq:weak_lensing_full} \ea where the indices
$i,j\equiv(\theta,\varphi)$ stand for the polar 2D coordinate
indices on the sphere, $\sigma_1$ and $\sigma_3$ are the Pauli
matrices defined on the 2D spherical sky, $g_{ij}$ is the 2D metric
on that surface given by $g=diag(1,{\sin^2 \theta})$ and $\phi$ is
the lensing potential related to the gravitational potential
$\Phi$ by Eq.~(\ref{eq:pot}). Here, and throughout, $\gamma_1$ and
$\gamma_2$ are the components of the weak lensing shear produced
by the gravitational tidal field, and which may be conveniently
written as a complex shear $\gamma(\rr) \equiv
\gamma_1(\rr)+i\gamma_2(\rr)$. They correspond to the two
orthogonal modes of the distortion which are, a priori, measurable
on the sky with respect to a chosen fixed coordinate system. At
fixed $r$, comparison of $\gamma$ at different locations
($\theta,\varphi$) on the sphere is not possible.  In curved
space, comparisons are only possible locally, and, as $\gamma$ is
not a scalar, parallel-transport of $\gamma$ from place to place
is path-dependent. This fact lies at the source of the interest in
first developing rotationally invariant CMB polarisation
components in terms of electric and magnetic
components~\cite{art:ZaldSeljak97} as we will see shortly in the
context of weak lensing. The differential operator $\nabla$
corresponds to a 2D covariant derivative on the sphere of radius
$r$, which for a scalar is given by the partial derivative
$\nabla_i = \partial_i$, for a covariant vector $X_i$ (i.e. rank-1
covariant tensor) is given by $\nabla_j X_i = \de_j X_i -
\Gamma^k_{ji}X_k$ and for a contravariant vector $X^i$ is given by
$\nabla_j X^i = \de_j X^i + \Gamma^i_{kj}X^k$ with the Christoffel
symbol $\Gamma^k_{ji}$ depending on the metric $g_{ij}$ (see for
instance~\cite{book:Peebles93}). In our case, for which
$g=diag(1,{\sin^2 \theta})$, we will have only three non-zero
Christoffel symbols which are 
$\Gamma_{11}^0  = -{\sin \theta} {\cos \theta}$ and 
$\Gamma_{01}^1  = \Gamma_{10}^1 = {\cot \theta}$. 
By construction, the
Christoffel symbol is always symmetric in its lower indices
$\Gamma^k_{ji} = \Gamma^k_{ij}$. Due to this symmetry property,
there is an important result for scalar fields $X$ which 
is that $\nabla_i \nabla_k X =\nabla_k \nabla_i X$. 
In the remaining of this work we will apply this property to the
lensing potential $\phi$.

The lensing potential $\phi$ can alternatively be observed by
means of the magnification via the isotropic convergence scalar
field  $\kappa$ defined as \be [\kappa\,(\rr)]_{ij} =
\kappa\,(\rr) I_{ij}=\frac{1}{2}g_{ij} \nabla^2 \phi (\rr)
\label{eq:conv_full} \ee where $\nabla^2=\nabla_i \nabla^i$ and
$I_{ij}$ is the Identity matrix. We name $[\kappa]_{ij}$ the
convergence field tensor. Note that the weak lensing regime
corresponds to lensing configurations such that $| \kappa| \ll 1$
and $|\gamma| \ll 1$. In this specific regime, the magnification
and the distortion of galaxy sources are so small that one cannot
measure them individually, rather one needs to perform a
statistical study of the lensed population.

It will be useful later to have the expressions~(\ref{eq:weak_lensing_full}) and
(\ref{eq:conv_full}) explicitly expanded in terms of covariant
derivatives on the 2D spherical sky. Transforming from the 
Cartesian $\{\ex(\nhatb),\ey(\nhatb) \}$ to a spherical
polar 2-dimensional coordinate system
$\{ \etheta(\nhatb),\ephi(\nhatb) \}$, where we choose
the basis of the two coordinate systems to be aligned
such that $dx=d\theta$ and $dy=\sin \theta d \varphi$, we can
explicitly express Eq.~(\ref{eq:weak_lensing_full}) for the weak
lensing shear as \ba [\gamma\,(\rr)\,]_{ij} & = & \left(
\begin{array}{cc}
        \gamma_1(\rr)                 &    {\sin \theta}\, \gamma_2(\rr)\\
        {\sin \theta}\,\gamma_2(\rr)  &   -  {\sin^2 \theta}\,\gamma_1(\rr)
\end{array} \right)  =
\left( \begin{array}{cc}
  \frac{1}{2}[\nablat\nablat - {\csc^2\theta} \nablap \nablap]
& \nablap\nablat\\
\nablap\nablat
&   \frac{1}{2}[\nablap\nablap - {\sin^2\theta} \nablat \nablat]
\end{array} \right) \,\phi(\rr).
\label{eq:weak_lensing_full_2} \ea
$[\gamma]_{ij}$ is the
$2\times2$ symmetric and traceless cosmic shear tensor field
associated to a source at a 3D position $\rr$ defined on the sky
and again $\nabla_i$ corresponds to a 2D covariant derivative on
the sphere. We used the general result for scalar fields $\nablat\nablap\,\phi =
\nablap\nablat\,\phi$. The $\sin\theta$
terms in the weak lensing shear tensor $[\gamma]_{ij}$ appear
because the orthogonal $(\theta,\varphi)$ basis is not
orthonormal. This result has been repeatedly used when describing
the polarisation tensor $\mathcal{P}_{ij}$ of the CMB on the
full-sky (see for example~\cite{art:KamionEtAl96}). One can
extract from Eq.~(\ref{eq:weak_lensing_full_2}) the individual
components of the weak lensing shear $\gamma_1$ and $\gamma_2$ \ba
\gamma_1(\rr) & = & \frac{1}{2}[\nablat\nablat -
                             {\csc^2\theta}\nablap \nablap]\,\phi(\rr) , \nn
\gamma_2(\rr) & = & \csc \theta \nablap\nablat \,\phi(\rr) .
\label{eq:gamma_12_full}
\ea
Similarly, one
obtains for the convergence field tensor $[\kappa]_{ij}$ given
by Eq.~(\ref{eq:conv_full}) the following expanded expression
\ba
[\kappa\,(\rr)\,]_{ij} & = &
\left( \begin{array}{cc}
        1    &    0 \\
        0    &   {\sin^2 \theta}
\end{array} \right) \kappa\,(\rr) =
\left( \begin{array}{cc}
  \frac{1}{2}[\nablat\nablat + {\csc^2\theta} \nablap \nablap]
& 0 \\
  0
& \frac{1}{2}[\nablap\nablap + {\sin^2\theta} \nablat \nablat]
\end{array} \right) \,\phi(\rr)
\label{eq:convergence_full_1}
\ea
where $\kappa$, the scalar convergence field in the 3D
full-sky, is given by
\ba
\kappa\,(\rr) =  \frac{1}{2}[\nablat\nablat + {\csc^2\theta} \nablap \nablap]\,\phi(\rr).
\label{eq:convergence_full_2}
\ea

\subsubsection{Weak lensing on the flat-sky}

When analysing data on a small patch of the sky, one can use the
flat-sky, or small-angle, approximation. There, one
defines the standard Cartesian coordinate system
$\{\ex(\nhatb),\ey(\nhatb) \}$ with metric given by $g=diag(1,1)$.
The covariant derivatives simplify to the standard partial
differentiation operators $\nabla_i \rightarrow \partial_i$.
The above Eq.~(\ref{eq:weak_lensing_full_2}) then
reduces to~\cite{art:CrittendenEtAl00}
\ba
[\gamma\, (\rr)\,]_{ij}& = &
\left( \begin{array}{cc}
        \gamma_1(\rr)  &  \gamma_2(\rr)\\
        \gamma_2(\rr)  & - \gamma_1(\rr)
\end{array} \right)
=  (\de_i \de_j - \frac{1}{2} \delta_{ij} \nabla^2) \,\phi (\rr)
\label{eq:weak_lensing_flat}
\ea
in such a way that we recover the well-known expressions for the components $\gamma_1$
and $\gamma_2$ of the shear on the 3D flat-sky
\ba
\gamma_1 (\rr) & = & \frac{1}{2}(\partial_x^2 - \partial_y^2 )\,\phi (\rr)\nn
\gamma_2 (\rr) & = & \partial_x\partial_y \,\phi (\rr)
\label{eq:shear_flat}
\ea
The convergence field takes the form
\be
\kappa (\rr)= \frac{1}{2}\nabla^2 \phi (\rr) =
\frac{1}{2}(\partial_x^2 + \partial_y^2) \phi (\rr)
\label{eq:conv_flat}
\ee
where $\nabla^2=\partial_i \partial^i=\partial_i \partial_i$ as the
metric is given by $g=diag(1,1)$.

\subsection{Weak lensing shear as a spin weight 2 object} 
\label{subsec:wl_edth}

Weak lensing induces a complex shear field $\gamma(\rr)$ which
transforms under a rotation (by an angle $\psi$ in the
anti-clockwise direction in our conventions) of the fixed
coordinate system according to $\gamma \rightarrow \gamma
e^{-is\psi}$, where $s=2$ is its {\em spin weight}. This phase
dependence expresses the fact that the complex shear field
$\gamma(\rr)$ is invariant under a rotation over $\pi$ radians.
Also, the two components of the shear field, $\gamma_1$ and
$\gamma_2$, are related by a $\pi/4$ radians rotation which
transforms one field into the other: $\gamma_1'=-\gamma_2$ and
$\gamma_2'=\gamma_1$ where the prime denotes the transformed
fields. In this subsection we use a description of the field based
on a geometrical differential operator called {\it edth} and
symbolised by $\edth$. This operator was firstly introduced by
Newman and Penrose in 1966~\cite{art:NewmanPenrose66} (see
also~\cite{art:GoldbergEtAl67}) on the surface of the sphere in
order to define the now widely used `spin-weight $s$ spherical
harmonics' which live on the 2D spherical space. A spin-weight $s$
spherical harmonic, symbolised as $\sYlm$, can be seen as a
generalisation of the scalar, vector and tensor spherical
harmonics (see~\cite{book:VarshalovichEtAl88,art:Challinor05}).
The $\sYlm$ (not defined for $|s| > l$) form a complete
orthonormal basis for each $s$, a result that we will use later in
order to describe the spin-weight $s$ functions on the 2D sphere.
The spin $s$ of a function is related to its transformation
properties under a rotation of the frame field where it is defined
(for example a scalar is a spin-0 function). The $\edth$ (and its
complex conjugate $\edthbar$) act as raising (and lowering)
operators on the 'quantum number' $s$ (integral number), such that
$\sYlm$ are eigenfunctions of $\edth\edthbar$ and can be obtained
by applying $\edth$ to the standard spin-weight $0$ spherical
harmonics $\Ylm$. The operator $\edth$ is effectively a covariant
differentiation operator acting on the surface of the sphere which
enables one to relate quantities of different spin. In particular,
it allows one to conveniently relate spin-weight $s$ objects,
which are not invariant under rotations of the coordinate frame,
to scalar quantities, which are invariant quantities under
rotations. All invariant differential operators on the sphere may
be expressed in terms of it.

The operator $\edth$ and the concept of spin-weight s functions
have been generalised to any 2D Riemannian manifold
(a 2D manifold with a metric) (e.g. ~\cite{art:LewisEtAl01,art:Norton Bartnick99}). 
For details and further
references see the Appendix, in particular Eq.~(\ref{eq:edth_sfunc_1})
for the expressions relating the $\edth$ operator to covariant
derivatives in a 2D Riemannian manifold.

As pointed out originally by Newman \&
Penrose~\cite{art:NewmanPenrose66}, any spin-weight $s$ function
defined on a 2D Riemannian manifold can be uniquely decomposed
into a scalar gradient (or electric/even) `E'-component and a
scalar curl (or magnetic/odd) `B'-component. We say that the
spin-weight $s$ field $\eta$ is even if $\eta = \edth^s f$ and odd
if $\eta = i \edthbar^s f$ for some real-valued spin-weight $0$
function $f$ (if $s < 0$ then we interpret $\edth^s$ as $(-
\edthbar)^{\mid s \mid}$). For $s = 1$ this decomposition
corresponds exactly to the classical Hodge-Helmholtz decomposition
of a vector field into the sum of a gradient and a curl component.

As a spin-weight 2 object, the shear field can be written as the
second edth-derivative of a complex
potential~\cite{art:NewmanPenrose66,art:KamionEtAl97,art:ZaldSeljak97,art:Stebbins96,art:KamionEtAl98,art:CrittendenEtAl00,art:SchneiderEtAl02}:
\begin{eqnarray}
\gamma   (\rr) \equiv  \gamma_1(\rr) +
i\gamma_2(\rr)
                 = {1\over 2}\edth \edth[\phi_E(\rr) + i\phi_B(\rr)],
\label{eq:gamma_decomp_1a}\\
\gamma^* (\rr) \equiv  \gamma_1(\rr) - i\gamma_2(\rr)
                 =  {1\over 2}\edthbar \edthbar[\phi_E(\rr) - i\phi_B(\rr)]
\label{eq:gamma_decomp_1b}
\end{eqnarray}
where we introduce two scalar real functions
$\phi_E(\rr)$ and $\phi_B(\rr)$ for the even and odd parts. The
normalisation factor of $2$ was chosen so that one can later identify
immediately the lensing potential to the even field $\phi_E(\rr)$.
As we will see, weak shear is
derivable from a real (lensing) potential $\phi_E(\rr)=\phi(\rr)$,
requiring $\phi_B(\rr)=0$.
This definition differs from the convention used in CMB
polarisation studies (see for instance~\cite{art:LewisEtAl01,art:BunnEtAl02})
and comes about due to the conventions adopted
for the spin-weight formalism (see the Appendix~\ref{appsubsec:spin_weight_Riem}).
Note that the scalar functions $\phi_E$ and $\phi_B$ introduced
are invariant under rotations of the reference frame.

We emphasise that the spin-raising and lowering operators $\edth$
and $\edthbar$ act on the 2D manifold at a distance $r$. We
are interested in their expressions both in the full-sky 2D sphere
and in the flat-sky 2D Euclidean space. Their derivation in both
geometries for any value of spin $s$ is detailed in the
Appendices~\ref{appsubsec:spin_weight_full} and ~\ref{appsubsec:spin_weight_flat}
and we give the final expressions
for $\edth$ and $\edthbar$ here. In the 2D spherical
full-sky we have
\ba \edth
\sfunc (\theta,\varphi) & = &   - (\de_{\theta}
                                    + i {\csc \theta}\de_{\varphi}
                                    - s {\cot \theta} )
                                     \sfunc (\theta,\varphi) \nn
                           & = &  - {\sin^s \theta}\,(\de_{\theta}
                                             + i{\csc\theta}\,\de_{\varphi})\,
                                    {\sin^{-s} \theta} \sfunc (\theta,\varphi),
\label{eq:edth_sfunc_sph}
\ea
and
\ba \edthbar
\sfunc (\theta,\varphi)  & = &    - (\de_{\theta}
                                    - i {\csc \theta}\de_{\varphi}
                                    + s {\cot \theta} )
                                     \sfunc (\theta,\varphi) \nn
                                & = &   - {\sin^{-s} \theta}\,(\de_{\theta}
                                           - i {\csc \theta}\,\de_{\varphi})\,
                                     {\sin^s \theta} \sfunc (\theta,\varphi).
\label{eq:edthbar_sfunc_sph}
\ea
In the 2D flat-sky the latter reduce to
\ba
\edth    \sfunc (x,y)& = & - (\de_x + i \de_y) \sfunc (x,y), \nn
\edthbar \sfunc (x,y)& = & - (\de_x - i \de_y) \sfunc (x,y).
\label{eq:edth_edthbar_flat}
\ea
If we use these results, or rather these results expressed in terms of covariant
derivatives, Eq.~(\ref{eq:gamma_decomp_1a}) can be re-expressed explicitly
in terms of covariant derivatives on the sphere and in the tensorial
formalism used in section~\ref{subsec:wl_tens} as
\ba
[\gamma\,(\rr)\,]_{ij}
 =
\left( \begin{array}{cc}
        \gamma_1(\rr)                 &    {\sin \theta}\, \gamma_2(\rr)\\
        {\sin \theta}\,\gamma_2(\rr)  &   -  {\sin^2 \theta}\,\gamma_1(\rr)
\end{array} \right)  & = &
\left( \begin{array}{cc}
    \frac{1}{2} [\nablat \nablat - {\csc^2 \theta} \,\nablap\nablap]
  & \nablap \nablat  \\
    \nablap \nablat
  & \frac{1}{2} [\nablap\nablap - {\sin^2\theta}  \,\nablat\nablat]
\end{array} \right) \,\phi_E(\rr) \label{eq:gamma_decomp_3} \\
 & + &
\left( \begin{array}{cc}
      -{\csc \theta}\,\nablap\nablat
  &   \frac{1}{2} [{\sin \theta}\,\nablat\nablat - {\csc \theta}\,\nablap\nablap]\\
      \frac{1}{2}[{\sin \theta}\,\nablat\nablat - {\csc \theta}\,\nablap\nablap]
  &    {\sin \theta}\,\nablap\nablat
\end{array} \right) \,\phi_B(\rr) \nonumber
\ea 
where we obeyed to the conventions
defined in the Appendices~\ref{appsubsec:spin_weight_Riem}
and~\ref{appsubsec:spin_weight_full}. In particular, we projected
the complex cosmic shear $\gamma$ into a tensor $[\gamma\,]_{ij}$
by means of relation Eq.~(\ref{eq:spin_tensor}) and again used the
general result $\nablat\nablap\,\phi = \nablap\nablat\,\phi$. We
can now compare directly the general E-B decomposition of the
shear Eq.~(\ref{eq:gamma_decomp_3}) with the theoretical
predictions for gravitational weak lensing shear on the 3D
full-sky Eq~(\ref{eq:weak_lensing_full_2}). We can do this because
the choice of the 2D polar coordinate system in both cases is the
same. It is then mathematically straightforward to identify
$\phi_E$ with the lensing potential $\phi$ such that
$\phi_E(\rr)=\phi(\rr)$ and $\phi_B(\rr) = 0$. Similarly one can
make such an identification in the flat-sky approximation and
re-express explicitly Eq.~(\ref{eq:gamma_decomp_1a}) in terms of
partial derivatives on a 2D Euclidean space in the same tensorial
formalism as above 
\ba 
[\gamma\,(\rr)\,]_{ij} & = &
   (\de_i \de_j - \frac{1}{2}\delta_{ij}\nabla^2)\phi_E(\rr)
   + \frac{1}{2}(\varepsilon_{kj}\de_i \de_k
   + \varepsilon_{ki} \de_k \de_j)\phi_B(\rr)
\label{eq:gamma_decomp_flat} 
\ea 
where $\varepsilon_{ij}$ is the
anti-symmetric 2-dimensional Levi-Civita tensor for which
$\varepsilon_{00}=\varepsilon_{11}=0$ and
$\varepsilon_{01}=-\varepsilon_{10}=1$~\cite{art:CrittendenEtAl00}.
If we compare this equation with the weak lensing theoretical
expression in the flat-sky limit Eq.~(\ref{eq:weak_lensing_flat}),
the identification between $\phi$ and $\phi_E$ is, as expected,
immediate. Therefore the shear field induced by gravitational
tidal fields only produces an E-pattern in cosmic shear maps. Such
a result has been shown
before~\cite{art:Kaiser92,art:Kaiser95,art:Stebbins96,art:KamionEtAl98,art:CrittendenEtAl00}
and has been applied to weak lensing data in the flat-sky
limit~\cite{art:WaerbekeEtAl01,art:HoekstraEtAl02,art:HamanaEtAl02,art:BrownEtAl03,art:JarvisEtAl03}.
It was to be expected as density (scalar) perturbations only
produce $E$-type effects. Hence the decomposition of the weak
lensing data into E and B components presents a clear advantage
over the $\gamma_1$ and $\gamma_2$ decomposition. On the full-sky
it allows one to isolate the effect caused by weak lensing and
disentangle it from any non-lensing contributions like noise,
foregrounds or systematics which should contribute similarly to
both the E and the B modes. It is worth noting that both
gravitational waves~\cite{art:Stebbins96}, multiple light lensing
scattering effects and source clustering~\cite{art:SchneiderEtAl02}
may also induce B modes, but their levels are expected to be
small. As we will mention shortly, one can have additional
spurious curl-type effects caused by the observational
pixelisation or by finite fields, due to a leakage between the E
and the B modes ~\cite{art:BunnEtAl02,art:LewisEtAl01}. In such a
situation the B-mode acts as a vital test for any
non-gravitational signal in the data. Aside from such
complications arising by finite fields and pixelisation, the E and
B potentials have also the advantage of being rotationally
invariant and so there are no ambiguities in their definition
related to the rotation of the coordinate system on the sphere.
Their interpretation is thus clearer than the one of the
$\gamma_1$ and $\gamma_2$ decomposition over the whole sky.

Having established this correspondence in 3D, one can recast both
Eq.~(\ref{eq:weak_lensing_full}) and
Eq.~(\ref{eq:weak_lensing_flat}), as the following valuable
relation between the weak lensing shear $\gamma$ and the lensing
potential in terms of the $\edth$ and $\edthbar$ operators \ba
\gamma   (\rr) = {1\over 2}\edth \edth \phi (\rr),
\label{eq:gamma_decomp_a}\\
\gamma^* (\rr) = {1\over 2}\edthbar\edthbar \phi (\rr)
\label{eq:gamma_decomp_b} \ea or, alternatively, the equivalent
relation for the two orthogonal components of $\gamma$
\begin{eqnarray}
 \gamma_1   (\rr) & = &  {1\over 4}(\edth \edth + \edthbar
\edthbar ) \phi (\rr), \nn \gamma_2   (\rr) & = & - {i\over 4}
(\edth \edth - \edthbar \edthbar ) \phi (\rr).
\label{eq:gamma_decomp_bis} \ea

Similar expressions can be obtained for the scalar convergence
field which is a spin-weight $0$ object.
As shown in the Eq.~(\ref{eq:laplacian}) of the Appendix, the Laplacian
acting on a rank-0 tensor is equivalent to applying the combination
$[\edth \edthbar + \edthbar \edth]/2$ to the corresponding spin-weight $0$
quantity. In particular, if one compares directly
Eq.~(\ref{eq:laplacian_fullsky}) expressed in the full-sky
to Eq.~(\ref{eq:convergence_full_2})
one sees that Eq.~(\ref{eq:convergence_full_2}) can equivalently be written
as
\ba
\kappa(\rr) & = &  \frac{1}{4}[\edth \edthbar + \edthbar \edth]\phi(\rr).
\label{eq:conv_decomp}
\ea
Such an identification can also be performed in the flat-sky
case. Note that there are natural extensions to higher spin-weight
objects: ${\cal F}(\rr)=-\frac{1}{6}(\edthbar \edth \edth +
\edth \edthbar \edth + \edth \edth \edthbar) \phi(\rr)$ and
${\cal G}(\rr)=-\frac{1}{2}\edth \edth \edth \phi(\rr)$ are the
analogues of the first and second flexion respectively
(see~\cite{art:GoldbergBacon04,art:BaconEtAl05}).

We now have both the weak lensing shear and convergence fields
expressed in the {\it edth} formalism which encompasses both the
full-sky and the flat-sky limits. Due to the role of the $\edth$
($\edthbar$) differential operators as spin-raising (lowering) the
spin of spin-weight $s$ spherical harmonics, this alternative
compact mathematical tool will be of particular beneficial use
when studying weak lensing in the 3D full-sky.


\section{Decomposition of 3D full-sky weak lensing into spin-weight spherical harmonics}
\label{sec:wl_full}

In the previous sections we have expressed the weak lensing shear
and convergence fields as spin-weight functions. Aided by the {\it
edth} formalism we have established their relation with the weak
lensing potential by means of a simple and unique E-B
decomposition scheme. In the present section, we concentrate on
the 3D (geometrically flat) spherical sky where the previous
fields can be decomposed into a combination of spin-weight
spherical harmonics and Bessel functions. We thereby derive the
relation between the expansion coefficients of the weak lensing
quantities and the expansion coefficients of the weak lensing
potential. The latter is readily related to the expansion
coefficients of the gravitational potential and hence to those of
the matter density field.

\subsection{Representation of a general distortion field in
spin-weight spherical harmonics}
\label{subsec:dist_sph_harm}

For spin-weight 2 fields such as the weak shear field, the natural
3D basis functions are products of radial functions and spin-weight 2
spherical harmonics.  With this choice, the 3D expansion coefficients
of $\gamma$ are related very simply to the expansion coefficients
of the lensing potential, where the latter is expanded in products
of ordinary (spin weight 0) spherical harmonics and the same
radial functions.

Let us first assume a general distortion shear field $\eta(\rr)$.
This field can be decomposed in two orthogonal components: $\eta
(\rr) \equiv \eta_1(\rr) + i\eta_2(\rr)$. Because of its
transformation properties, such a field is a spin-weight $2$
object. By consequent it can be decomposed into an even/`E'
and an odd/`B' part by means of two scalar real functions defined
on the 2D sphere of radius $r$, $\phi_E$ (for the even part) and
$\phi_B$ (for the odd part) as in Eqs.~(\ref{eq:gamma_decomp_1a}) 
and (\ref{eq:gamma_decomp_1b})
\begin{eqnarray}
\eta   (\rr) = {1\over 2}\edth \edth[\phi_E(\rr) + i\phi_B(\rr)]
&\, ,\, & \eta^* (\rr) = {1\over 2}\edthbar \edthbar[\phi_E(\rr) -
i\phi_B(\rr)]. \label{eq:eta_decomp}
\end{eqnarray}
As discussed previously, the two real scalar functions $\phi_E$
and $\phi_B$ introduced completely characterise the distortion
field. They have the advantage of being scalars which are
invariant under any rotation of the coordinate system. We can then
expand them in 3D on the full sky as we expanded the lensing
potential field in Eq.~(\ref{eq:general_3D_expansion}) in terms of
the standard (spin-0) spherical harmonics $\Ylm$ and of Bessel
functions
\begin{eqnarray}
\phi_E   (\rr) & = & - 2 \int_{0}^{\infty}d k \sum_{\ell=0}^\infty \sum_{m=-\ell}^\ell {\sqrt
                      \frac{(\ell-2)!}{(\ell+2)!}}
                      \,a_{E,\lm}(k) \Zklm(r,\theta,\varphi),\nn
\phi_B   (\rr) & = & - 2 \int_{0}^{\infty}d k\sum_{\ell=0}^\infty \sum_{m=-\ell}^\ell {\sqrt
                      \frac{(\ell-2)!}{(\ell+2)!}}
                      \,a_{B,\lm}(k) \Zklm(r,\theta,\varphi)
\label{eq:e_b_pot_decomp}
\end{eqnarray}
where we introduced the orthonormal basis functions $\Zklm$ for simplicity and
which, in a spatially flat background geometry, are given by
\begin{equation}
\Zklm(r,\theta,\varphi) = {\sqrt \frac{2}{\pi}}\, k \, j_{\ell}(kr)\,\Ylm
                    (\theta,\varphi).
\label{eq:zklm_def}
\end{equation}
Apart from the normalisation factor of 2, the normalisation terms
and the sign in Eqs.~(\ref{eq:e_b_pot_decomp}) were introduced to
maintain consistency with previous works on CMB
polarisation~\cite{art:BunnEtAl02}. The factor $2$ is linked to
the choice of the spin-weight conventions as established in the
Appendix~\ref{appsec:spin_weight} and is necessary if one wants to
recover relations between the expansion coefficients of $\gamma$
and of the fields $\phi_E$ and $\phi_B$ which are analogous to the
relations between the CMB polarisation field $\mathcal{P}_{ij}$
and the electric and magnetic scalar fields~\cite{art:BunnEtAl02}.
We point out here that we do not assume 3D homogeneity and
isotropy (at a fixed instant in time) for the scalar fields
$\phi_E$ and $\phi_B$ and so do not introduce the time $r$
dependence in the coefficients of their expansion as discussed in
section~\ref{subsec:grav_pot}. If we now substitute
Eqs.~(\ref{eq:e_b_pot_decomp}) into Eqs.~(\ref{eq:eta_decomp}) and
use the properties of the operators $\edth$ and $\edthbar$ acting
on the standard (spin-weight 0) spherical harmonics (see Appendix
Eqs.~(\ref{eq:Prop_edth_sYlm}))
one obtains naturally the following relations for $\eta$ 
\ba 
\eta
(\rr) & = &\int_{0}^{\infty}d k \sum_{\ell=0}^\infty
\sum_{m=-\ell}^\ell\,_{\,_2}\!\eta_{\lm}(k)
                                          \ZklmpII(r,\theta,\varphi), \nn
\eta^* (\rr) & = &\int_{0}^{\infty}d k \sum_{\ell=0}^\infty \sum_{m=-\ell}^\ell\,_{\,_{-2}}\!\eta_{\lm}(k)
                                      \ZklmmII(r,\theta,\varphi)
\label{eq:eta_expansion_sylm}
\ea
where the expansion coefficients are related to the
$\phi_E$ and $\phi_B$ expansion coefficients by~\cite{art:ZaldSeljak97,art:BunnEtAl02}
\ba
     \,_{\,_2}\!\eta_{\lm} (k)    =  -[a_{E,\lm} +ia_{B,\lm}] (k)
& , & \,_{\,_{-2}}\!\eta_{\lm}(k) =  -[a_{E,\lm} -ia_{B,\lm}] (k);
\label{eq:e_b_eta_decomp_1} \\
           a_{E,\lm}(k)  =  -{1\over 2}[\,_{\,_2}\!\eta_{\lm} + \,_{\,_{-2}}\!\eta_{\lm}](k)
& , &      a_{B,\lm}(k)  =  {i\over 2}[\,_{\,_2}\!\eta_{\lm} -
\,_{\,_{-2}}\!\eta_{\lm}](k). \label{eq:e_b_eta_decomp_2} \ea and
where now the orthonormal basis functions $\ZklmpmII$ are expressed in terms
of a set of functions, the spin-weight $\pm 2$ spherical harmonics
$\pmtYlm$, the ideal basis to express any spin-weight $\pm2$
function 
\ba 
\sZklm(r,\theta,\varphi) & = & {\sqrt
\frac{2}{\pi}}\, k \, j_{\ell}(kr)\,\sYlm (\theta,\varphi) 
\label{eq:basis_func_3D} 
\ea
where the spin is given by $s=\pm2$.  
As we can see, any distortion field $\eta$ defined on the 3D
full-sky is most naturally expressed in terms of spin-weight $2$
spherical harmonics, which define a set of orthonormal basis on
the surface of the sky. The {\it edth} formalism introduced in the
previous section has just shown its mathematical advantages when
working in spherical space with spherical harmonics.

Inserting Eqs.~(\ref{eq:e_b_eta_decomp_1}) into
Eqs.~(\ref{eq:eta_expansion_sylm}) we can also relate the $\eta_1 $
and $\eta_2$ orthogonal components to the E-B decomposition expansion
coefficients of the distortion field as
\begin{eqnarray}
\eta_1
(\rr) & = &
            -\int_{0}^{\infty}d k  \sum_{\ell=0}^\infty \sum_{m=-\ell}^\ell
             [a_{E,\lm}(k)X_{1,k\lm}(r,\theta,\varphi)+ia_{B,\lm}(k)X_{2,k\lm}(r,\theta,\varphi)], \nn
\eta_2 (\rr) & =  &
            -\int_{0}^{\infty}d k  \sum_{\ell=0}^\infty \sum_{m=-\ell}^\ell
             [a_{B,\lm}(k) X_{1,k\lm}(r,\theta,\varphi)-ia_{E,\lm}(k)X_{2,k\lm}(r,\theta,\varphi)]
\label{eq:e_b_gamma_decomp_2}
\end{eqnarray}
where we defined two new basis sets: $X_{1,k\lm}=(\ZklmpII +
\ZklmmII)/2$ and $X_{2,k\lm}=(\ZklmpII - \ZklmmII)/2$. As we have
the relations $X^*_{1,k\lm}=-X_{1,k-\lm}$,
$X^*_{2,k\lm}=-X_{2,k-\lm}$, $a^*_{E,\lm}=a_{E,-\lm}$ and
$a^*_{B,\lm}=a_{B,-\lm}$, the coefficients $\gamma_1$ and
$\gamma_2$ are real quantities.

Both representations can be chosen to study a general observed
distortion field $\eta$, i.e. either use the $\eta_1$ and $\eta_2$
components or transform them into $E$ and $B$ type quantities.
They contain the same information but, as explained before, the
$E$ and $B$ are rotationally invariant on the full-sky and behave
differently under parity transformation
~\cite{art:NewmanPenrose66,art:Hu00}. In terms of the expansion
coefficients, $a_{E,\lm}(k)$ has parity $(-1)^{\ell}$ while
$a_{B,\lm}(k)$ has parity $(-1)^{\ell+1}$, a property which may
certainly be useful when trying to characterise the origin of the
distortion field observed on the sky. In particular, the E-B
decomposition components have a special meaning in the context of
weak lensing studies as weak lensing is in nature an E-type field
and does not produce any B-type component, as we saw before in
section~\ref{subsec:wl_edth}.

In practice, if one wants to determine the real scalar functions
$\phi_E$ and $\phi_B$ describing a given distortion field on the
sky one should use Eqs.~(\ref{eq:eta_decomp}). This is a trivial
task if our observations are performed over the full-sky because
one can use the orthogonality of the spin-weight spherical
harmonics to obtain the E and B expansion coefficients from the
measured field $\gamma$. But full-sky observations are
unrealistic. Even if an experiment does cover the whole sky,
certain regions always need to be removed in order to minimise the
foreground contribution (such as bright stars). The perverse side
effect of partial sky-coverage is to induce an E-B mode mixing. In
this case the non-local decomposition~(\ref{eq:eta_decomp}) is not
unique. If one tries to solve for $\phi_E$ or $\phi_B$ by taking
linear combinations of second derivatives of the distortion field
one gets \ba \nabla^2(\nabla^2 +2) \phi_E & = & \frac{1}{2}
         [\edthbar\edthbar \eta + \edth\edth \eta^*], \nn
\nabla^2(\nabla^2 +2) \phi_B & = & \frac{i}{2}
        [\edthbar\edthbar \eta - \edth\edth \eta^*].
\ea This result enables one to use distortion observations to
construct the E and B distortion shear modes on the full-sky. But
over a cut-sky, solving this system implies the specification of
unknown boundary conditions (i.e. value of the potentials and
their derivatives at the boundaries). This problem has been
extensively studied by many in the context of CMB polarisation
measurements~\cite{art:LewisEtAl01,art:BunnEtAl02} but to date
separation of E-B modes has been performed statistically during
the estimation of the power spectra using the so-called
pseudo-$C_{\ell}$ method (see~\cite{art:HivonEtAl01}). On the
flat-sky and in the context of weak lensing, the 
locally defined aperture mass $M_{ap}$ may similarly provide
unambiguous
mode separation~\cite{art:KaiserEtAl94,art:SchneiderEtAl98}. A
similar statistic can be built for the weak lensing B mode. It is
beyond the scope of this work to enter into technical details
related to mode separation, so we refer the reader to
Challinor~\cite{art:Challinor05} for a useful recent CMB
polarisation review and van Waerbeke and
Mellier~\cite{art:WaerbekeMellier03} for an updated weak lensing
review.

\subsection{Representation of the weak lensing shear and convergence fields in
spin-weight spherical harmonics}
\label{subsec:wl_sph_harm}

The previous results can easily be applied to the case of the weak
lensing shear field $\gamma$ for which (see
Eqs.~(\ref{eq:gamma_decomp_a}) and~(\ref{eq:gamma_decomp_b})) \bas
\gamma   (\rr) = {1\over 2}\edth \edth \phi (\rr) &\, ,\, &
\gamma^* (\rr) ={1\over 2}\edthbar \edthbar \phi (\rr) \eas where
$\phi$ is the lensing potential defined in Eq.~(\ref{eq:pot}) and
which we can expand in 3D as in Eq.~(\ref{eq:general_3D_expansion})
where the expansion coefficients are $\phi_{\lm}(k)$.
Comparison of the 3D expansion of the lensing potential $\phi$ 
with Eq.~(\ref{eq:e_b_pot_decomp}) shows that the $a_{E,\ell m}$ and the
$\phi_{\lm}$ are related by 
\ba a_{E,\lm}(k) = -
\frac{1}{2}\sqrt{\frac{(\ell+2)!}{(\ell-2)!}}
                \phi_{\ell m}(k).
\label{eq:E_coefficients} 
\ea 
As demonstrated before, for
gravitational weak lensing, $a_{B,\lm}(k)=0$. The 3D full-sky
coefficients of the spin-weight $2$ shear field $\gamma$,
$\,_{\,_2}\!\gamma_{\ell m} (k)$ 
and $\,_{\,_{-2}}\!\gamma_{\ell m} (k)$, 
defined similarly to Eq.~(\ref{eq:eta_expansion_sylm})
are then related to $\phi_{\lm}$ by (see~\cite{art:Hu00,art:Taylor01,art:Heavens03})
\ba
\,_{\,_2}\!\gamma_{\ell m} (k) = \,_{\,_{-2}}\!\gamma_{\ell m} (k)
             =  \frac{1}{2}\sqrt{\frac{(\ell+2)!}{(\ell-2)!}} \phi_{\ell m}(k).
\label{eq:gamma_pot_lm} \ea The standard components of the weak
lensing shear $\gamma_1$ and $\gamma_2$ can be directly obtained
from this last relation. If one uses the
Eq.~(\ref{eq:e_b_gamma_decomp_2}) with $a_{B,\ell m}(k)=0$ and the
following expansion in 3D for the shear components $\gamma_1$ and
$\gamma_2$ 
\ba \gamma_1 (\rr) & = & \int_{0}^{\infty}d k
\sum_{\ell=0}^\infty \sum_{m=-\ell}^\ell
                     \gamma_{1,\lm}(k)X_{1,k\lm}(r,\theta,\varphi),\nn
\gamma_2 (\rr) & = & \int_{0}^{\infty}d k  \sum_{\ell=0}^\infty \sum_{m=-\ell}^\ell
                     \gamma_{2,\lm}(k)X_{2,k\lm}(r,\theta,\varphi)
\label{eq:gamma12_decomp}
\ea
where $X_{1,k\lm}$ and $X_{2,k\lm}$ were defined as in 
Eq.~(\ref{eq:e_b_gamma_decomp_2}), one immediately has
\ba
\gamma_{1,\ell m} (k) & = & i\,\gamma_{2,\ell m} (k) = - a_{E,\lm}(k),
\label{eq:gamma12_coefficients}
\ea
where the expression for $a_{E,\lm}$ in terms of the lensing potential
was defined in Eq.~(\ref{eq:E_coefficients}). As we see,
aside from the physical and experimental advantages
of the $E-B$ decomposition scheme, using either
of the expansion coefficients $\gamma_{1/2,\ell m}$ or $a_{E,\lm}$ is
equivalent as they are very simply related.

Likewise, we can do the same exercise for the convergence field $\kappa$.
From Eq.~(\ref{eq:conv_decomp}) and using
Eqs.~(\ref{eq:Prop_edth_sYlm}), the (scalar) expansion coefficients 
of $\kappa$ are given by~\cite{art:Hu00}
\ba
\kappa_{\ell m} (k) & = & - \frac{\ell(\ell + 1)}{2} \phi_{\ell m} (k).
\label{eq:kappa_coefficients}
\ea

To summarise, the coefficients of the expansion of the shear field
in spin-weight 2 spherical harmonics and spherical Bessel
functions, $\,_{\,_{\pm2}}\!\gamma_{\ell m} (k)$ are related to
those of the lensing potential $\phi_{\ell m}(k)$ by
Eq.~(\ref{eq:gamma_pot_lm}). The $\phi_{\ell m}(k)$ are related to
the gravitational potential $\Phi_{\ell m}(k)$ by
Eq.~(\ref{eq:transf_lens_grav}), and these are in turn related to
the overdensity field by Poisson's equation
(\ref{eq:poisson_bis}). In this way we establish a connection
between the observable shear quantities
$\,_{\,_{\pm2}}\!\gamma_{\ell m} (k)$ (and associated quantities)
and the overdensity field, whose statistical quantities have known
dependence on cosmological parameters. In the knowledge of such
results one may proceed to the derivation the 3D full-sky power
spectra (or any any relevant statistics) of the weak lensing shear
and convergence fields.


\section{Weak lensing 3D power spectra}
\label{sec:wl_power_spectra}

We shall now be interested in calculating the weak lensing shear
$\gamma$ and convergence $\kappa$ 3D power spectra and in relating
them to the 3D gravitational potential power spectra
$C^{\Phi\Phi}$ both in the full-sky and flat-sky approximation.

\subsection{Weak lensing 3D power spectra on the full-sky}
\label{subsec:wl_power_spectra_full}

If one performs a 3D spectral decomposition in the full-sky of a
statistically homogeneous and isotropic field $f(\rr;r)$ at time
instant defined by $r$ (one can use the comoving distance $r$ as a
measure of real time $t$ as
$r$ and $t$ are physically equivalent being related via
the scale factor $a$) with
coefficients $f_{\ell m } (k;r)$ as in
Eq.~(\ref{eq:general_3D_expansion_lm}), then the 3D power
spectrum of the field at time $r$, $C_{\ell}(k;r)$, is defined by
\ba 
\lgl f_{\ell m } (k;r)
f^*_{\ell' m' } (k';r) \rgl =
             C_{\ell}(k;r) \delta_D(k-k') \delta_{\ell \ell'}
            \delta_{mm'}
\label{eq:ps_general_3d_homog_isot} 
\ea 
where $\delta_{ij}$ is a
Kronecker delta function and $\delta_D(x)$ a 1D Dirac delta
function. The time dependence (introduced through $r$) of the
coefficients, at which the spectral expansion is performed, was
explained in section~\ref{subsec:grav_pot} and is necessary to
ensure the 3D spatial homogeneity (and isotropy) of the field we
are expanding at a certain time instant. Most importantly,
$C_{\ell}(k;r)$ is independent of $\ell$, and is simply the 3D
power spectrum $P(k;r)$:
\begin{eqnarray}
C_{\ell}(k;r) & = & P(k;r) \label{eq:cl_p_3d_homog_isot}
\end{eqnarray}
where $P(k;r)$ is obtained from performing a
3D Fourier expansion of the field $f(\rr;r)$ so that $\lgl
f(\kk;r) f^*(\kk';r)\rgl = (2\pi)^3 P(k;r) \delta_D^3(\kk-\kk')$.
Indeed the spherical expansion in 3D of
Eq.~(\ref{eq:general_3D_expansion}) for $f(\rr;r)$ is equivalent
to the 3D Fourier expansion \ba f(\rr;r) & =
&\frac{1}{(2\pi)^3}\int d^3\kk f(\kk;r)\, e^{i\kk.\rr}
\label{eq:3d_fourier_exp} 
\ea 
because of the well-known
identity~\cite{book:LiddleLyth00} 
\ba e^{i\kk.\rr} & = & 4\pi
\sum_{\ell}\sum_{m=-\ell}^{m=\ell}
            i^{\ell}j_{\ell}(kr) \Ylm (\theta_k,\varphi_k)\Ylm
                        (\theta_r,\varphi_r).
\label{eq:plane_wave_3d} \ea One can see this by substituting
Eq.~(\ref{eq:plane_wave_3d}) into Eq.~(\ref{eq:3d_fourier_exp})
and using the identity $d^3k= k^2 d\Omega_k dk$ where
$\Omega_k\equiv(\theta_k,\varphi_k)$ so that we obtain the
following relation between the expansion coefficients $f_{\ell m }
(k;r)$ (see Eq.~(\ref{eq:general_3D_expansion_lm})) and
$f(\kk;r)$~\cite{book:LiddleLyth00} \ba f_{\ell m } (k;r) =
\frac{1}{\sqrt{8\pi^3}}ki^{\ell}\int d\Omega_k
f(\kk;r)\Ylm(\theta_k,\varphi_k).\nonumber \ea One can then
substitute this last expression in
Eq.~(\ref{eq:ps_general_3d_homog_isot}) and use $\lgl f(\kk;r)
f^*(\kk';r)\rgl = (2\pi)^3 P(k;r) \delta_D^3(\kk-\kk')$. The trick
then relies in replacing $\delta_D^3(\kk-\kk')=\int d^3x
e^{i(\kk-\kk').\x}/(2\pi)^3$ and using the
identity~(\ref{eq:plane_wave_3d}). The orthogonality relations of
the spherical harmonics~(\ref{eq:sph_harm_orth_comp_1}) and of the
Bessel functions~(\ref{eq:orth_bessel}) as well as
$\Ylm^*=(-1)^mY_{\ell -m}$ will then be useful to reach the final
result, i.e. $C_{\ell}(k;r) =  P(k;r)$.

Now if we use the expansion coefficients of the
field located at different distances (or equivalently at different
times from us) say at $r$ and $r'$,
to effectively calculate the cross-power spectrum
of two different homogeneous and isotropic
fields, $f(\rr;r)$ and $f(\rr';r')$, then one can still argue that
the 3D homogeneity and isotropy argument holds:
\ba
\lgl a_{\ell m } (k;r) a^*_{\ell' m' } (k';r') \rgl =
             C_{\ell}(k;r,r') \delta_D(k-k') \delta_{\ell \ell'}
             \delta_{mm'}
\label{eq:cl_dif_r} \ea where now we have the identity
\begin{eqnarray}
 C_{\ell}(k;r,r') = P(k;r,r')
\label{eq:cl_p_3d_homog_isot_rr'}
\end{eqnarray}
where again $P(k;r,r')$ is defined by $\lgl f(\kk;r)
f^*(\kk';r')\rgl = (2\pi)^3 P(k;r,r') \delta_D^3(\kk-\kk')$. A
similar result is frequently utilised to determine the cross-power
spectra of different homogeneous and isotropic fields, like for
instance the weak gravitational lensing of the CMB and the cosmic
shear fields~\cite{art:Hu01}. It will be necessary to us when
calculating the power spectra of the gravitational potential field
$\Phi$ which comes about when calculating the spectra of the
lensing potential. Again we stress that, contrary to the
gravitational potential, the 3D lensing potential $\phi$ is not
homogeneous and isotropic in 3D space. It is given by a 2D
projection at each source distance $r$ of the gravitational
potential existing between us and the source and so it maintains
the homogeneity and isotropy characteristics of the gravitational
potential field on the 2D sky, but not in the radial direction. We
will then use for the field $\phi$ the relation
\begin{equation}
\lgl \phi_{\ell m } (k) \phi^*_{\ell' m' } (k') \rgl =
C^{\phi\phi}_{\ell}(k,k')\delta_{\ell \ell'}\delta_{mm'},
\end{equation}
where $C_\ell^{\phi\phi}$ is the 3D lensing potential power
spectrum. Contrary to the case of a 3D homogeneous and isotropic
field, an equivalent relation to Eq.~(\ref{eq:cl_p_3d_homog_isot})
does not hold any longer. Naturally, the lensing potential, the
shear and the convergence all share the same statistical
properties.

For completeness, although redundant, we present
the various possible shear power spectra
depending on the type of shear decomposition chosen.
If one uses the expansion coefficients of the shear $\gamma(\rr)$
itself, i.e.$\,_{\,_{\pm2}}\!\gamma_{\lm}(k;r)$ of
Eq.~(\ref{eq:gamma_pot_lm}),
then one gets
\ba
C^{\gamma \gamma}_{\ell}(k_1,k_2)& = &\frac{1}{4}\frac{(\ell+2)!}{(\ell-2)!} C^{\phi
                                \phi}_{\ell}(k_1,k_2)
\label{eq:PS_gammagamma} 
\ea 
where $C_\ell^{\phi\phi}$ is the
lensing potential power spectra. All the remaining power spectra
can then be conveniently expressed in terms of $C^{\gamma
\gamma}_{\ell}(k_1,k_2)$. For the $E$-$B$ decomposition
components, i.e. $a_{E,\ell m}(k)$ of
Eq.~(\ref{eq:E_coefficients}) and $a_{B,\ell m}(k)$, one
has~\cite{art:Hu00} 
\ba 
C^{E E}_{\ell}(k_1,k_2) & = & C^{\gamma
\gamma}_{\ell}(k_1,k_2),\nn C^{B B}_{\ell}(k_1,k_2) & = & 0.
\label{eq:PS_ee} 
\ea 
And finally for the $\gamma_1$ and $\gamma_2$
shear decomposition Eqs.~(\ref{eq:gamma12_decomp}) and
(\ref{eq:gamma12_coefficients}) one obtains 
\ba
C^{\gamma_1 \gamma_1}_{\ell}(k_1,k_2) & = & C^{\gamma_2
\gamma_2}_{\ell}(k_1,k_2) = C^{\gamma \gamma}_{\ell}(k_1,k_2),
\label{eq:PS_gamma1gamma2} 
\ea 
where for symmetry reasons $C^{\gamma_1 \gamma_2}=0$. 
Note that the expansions of $\gamma_{1}$ and $\gamma_{2}$ are in terms
of the orthogonal but not orthonormal functions $X_{1,k\lm}$ and $X_{2,k\lm}$, 
so $\gamma_{1}$ and $\gamma_{2}$ each contribute half to the power of 
$\gamma$, as expected. Concerning the
convergence field we have~\cite{art:Hu00} \ba C^{\kappa
\kappa}_{\ell}(k_1,k_2) & = & \frac{\ell^2(\ell+1)^2}{4} C^{\phi
                           \phi}_{\ell}(k_1,k_2).
\label{eq:PS_conv}
\ea

The next and final step is to relate the lensing and the
gravitational potential power spectra, $C_\ell^{\phi\phi}$ and
$C_\ell^{\Phi\Phi}$ respectively, by means of
Eq.~(\ref{eq:transf_lens_grav}). After some straightforward
algebra we obtain 
\ba 
C^{\phi\phi}_{\ell}(k_1,k_2) =
\frac{16}{\pi^2c^4}
                                \int_{0}^{\infty} k^2 \,dk
                \,I_\ell(k_1,k)I_\ell(k_2,k)
\label{eq:cl_phi}
\ea 
with
\begin{eqnarray}
\label{eq:Ieq}
I_\ell(k_i,k) \equiv k_i \int_{0}^{\infty}dr\,r\, j_{\ell}(k_ir)
               \int_{0}^{r}dr'\,\left[ \frac{r-r'}{r'}\right]
                  j_{\ell}(k
                  r')\sqrt{P^{\Phi\Phi}(k;r')}\label{eq:I}
\end{eqnarray}
where we have used Eq.~(\ref{eq:cl_dif_r}) applied to the
gravitational potential field and introduced the familiar 3D power
spectrum of the gravitational potential $P^{\Phi\Phi}$ via
Eq.~(\ref{eq:cl_p_3d_homog_isot_rr'}). For a non-uniform clustering of
sources, the number density needs to be introduced into 
the outer integral~\cite{art:HeavensKitching05}.
The correlations in the
potential field are significantly non-zero for small separations
(much smaller than the speed of light times the timescale over
which it evolves), so we have assumed for convenience that the
gravitational potential power spectrum can be accurately
approximated by
\begin{eqnarray}
P^{\Phi\Phi}(k;r,r') \simeq
\sqrt{P^{\Phi\Phi}(k;r)\,P^{\Phi\Phi}(k;r')}.
\end{eqnarray}
This can be seen by noting that Poisson's equation implies that
\begin{eqnarray}
\Phi(\kk;r) = - \frac{3\Omega_m H_0^2}{2a(t)k^2}\,\delta(\kk;r)
\end{eqnarray}
and so
\begin{eqnarray}
\lgl \Phi(\kk;r) \Phi^*(\kk';r') \rgl = 
\left( \frac{3\Omega_m H_0^2}{2} \right)^2
\int \frac{d^3 \rr d^3 \rr'}{a(t)a(t')}
\frac{\lgl \delta(\kk;r) \delta^*(\kk';r')  \rgl }{k^2 k'^2}
e^{-i\kk.\rr + i \kk'.\rr'}
\end{eqnarray}
and the correlation of $\delta$ is restricted to small scales 
$\mid \rr-\rr'\mid \leq 100$Mpc. The lookback time over such a distance
is small, so we can approximate $\rr \simeq \rr'$ (or $t \simeq
t'$). Therefore we can replace the power spectra
$P^{\Phi\Phi}(k;r,r')$ by either $P^{\Phi\Phi}(k;r)$ 
or $P^{\Phi\Phi}(k;r')$. For algebraic convenience, 
we choose the geometric mean of the power
spectra, which allows us to separate two internal integrals and
reduce computation time significantly. A further justification is apparent in 
Fig.~(\ref{fig:shearpowerdiag}): the Bessel functions cut off
long-wavelength contributions with $k \leq \ell/r_{max}$ where $r_{max}$
is the extent of the survey.
Practical issues of implementation are briefly discussed at the
end of the next section.  Note that the two-point statistics of
the shear field are dependent on the nonlinear potential power
spectrum, which is related to the nonlinear matter power spectrum
via Poisson's equation.  Accurate fits for the nonlinear matter
power spectrum are available to enough
wavenumbers~\cite{art:Smithetal03}, but the accuracy for
high-precision determination of dark energy properties needs to be
tested with simulations.

\subsection{Weak lensing 3D power spectra in the flat-sky limit}
\label{subsec:wl_power_spectra_flat}

In the flat-sky, we can expand a 3D field $f$ at 3D position $\rr
\equiv (r,\vth)$ on the sky into a combination of 2D Fourier modes
and Bessel functions in the radial direction $r$ 
\ba 
f (r,\vth) &
= &\sqrt{\frac{2}{\pi}}
                 \int_{0}^{\infty} k \, dk
                 \int_{0}^{\infty} \frac{d^2 \vell}{(2\pi)^2}
         f (k,\vell) j_{\ell}(kr) e^{i\vell.\vth},
\label{eq:general_3D_expansion_flat_1}\\
f (k,\vell)   & = &\sqrt{\frac{2}{\pi}}
         \int_{0}^{\infty} r^2 dr
         \int_{0}^{\infty} d^2 \theta
         f (r,\vth) k j_{\ell}(kr)
         e^{-i\vell.\vth}
\label{eq:general_3D_expansion_flat_2}
\ea
where the notation
$\vec{x}$ describes 2D vectors. Such an expansion was chosen to
maintain a direct relation with the 3D full-sky expansion
introduced previously. It thus presents the same attractive
advantages as its full-sky counterpart
(see discussion in section~\ref{subsec:grav_pot}).
One can now easily obtain the full-sky to flat-sky
correspondence, i.e. establish a relation between the expansion
coefficients $f (k,\vell)$ in the flat-sky as defined above and
$f_{\ell m} (k)$ in the full-sky defined in
Eq.~(\ref{eq:general_3D_expansion}).

For small angles around the pole of spherical coordinates, defined
by the angles $(\theta,\varphi)$ ignoring the radial dependence,
for which $\theta \rightarrow 0$, one can use the following
expansion of the plane 2D wave~\cite{art:Hu00,art:SantosEtAl02}
\be
 e^{i\vell.\vth} \simeq \sqrt{\frac{2 \pi}{\ell}}
            \sum_{m} i^m \Ylm(\theta,\varphi)
            e^{-im\varphi_{\ell}}
\label{eq:plane_wave_2D} \ee where $\vell=(\ell
\cos{\varphi_{\ell}},\ell \sin{\varphi_{\ell}})$ and $\vth=(\theta
\cos{\varphi},\theta \sin{\varphi})$. The vector $\vell$ can be
interpreted as the continuous limit of the integer that labels the
spherical harmonics $\Ylm$. Replacing Eq.~(\ref{eq:plane_wave_2D})
into Eq.~(\ref{eq:general_3D_expansion_flat_1}) and making use of
$\int_{0}^{\infty}d^2 \vell = \int_{0}^{\infty}\ell d\ell
\int_{0}^{2\pi}d\varphi_{\ell} \rightarrow \sum_{l} \ell
\int_{0}^{2\pi}d\varphi_{\ell}$ for high $\ell$, one obtains the
following relation between the 3D flat-sky and the 3D full-sky
coefficients 
\be f_{\ell m} (k) = \sqrt{\frac{\ell}{2 \pi}} i^m
         \int_{0}^{2\pi} \frac{d\varphi_{\ell}}{2\pi}
         e^{-im\varphi_l} f (k,\vell).
\label{eq:3D_full_flat_coef}
\ee
The inverse relation is
\be
f (k,\vell) = \sqrt{\frac{2 \pi}{\ell}}
                \sum_{m} i^{-m}
        f_{\ell m} (k) e^{im\varphi_l}.
\label{eq:3D_flat_full_coef}
\ee
We point out that, as expected, we recover the same relations
that were obtained by others for the 2D
case~\cite{art:Hu00,art:SantosEtAl02}, as the small-angle
limit only affects the 2D angular expansion, not the radial direction. As
the reader may have guessed, we also recover that the flat-sky
$\mathcal {C}^{\Phi\Phi}$ and the
full-sky $C^{\Phi\Phi}$ power spectra are
equivalent in the high-$\ell$ limit.
For a fully 3D homogeneous and isotropic field, like the gravitational
potential we have
\ba
\mathcal {C}^{\Phi\Phi} (k,l;r)
               & \stackrel{\ell \rightarrow \infty}{=} &
                C^{\Phi\Phi}_{\ell}(k;r) = P^{\Phi\Phi}(k;r)
\label{eq:3D_flat_full_grav} \ea where we have used
Eq.~(\ref{eq:ps_general_3d_homog_isot}), the definition $\lgl \Phi
(k,\vell) \Phi^* (k',\vell') \rgl = (2\pi)^2 \mathcal
{C}^{\Phi\Phi}(k,\ell)\delta_D(k-k')\delta_D^2(\vell-\vell')$ and
Eq.~(\ref{eq:cl_p_3d_homog_isot}). Again $P^{\Phi\Phi}$ is the 3D
gravitational potential power spectrum. For a 3D field which is
homogeneous and isotropic only on the 2D sky (e.g. the lensing
potential) \ba \mathcal {C}^{\phi\phi} (k_1,k_2,l)
       & \stackrel{\ell \rightarrow \infty}{=} &
               C^{\phi\phi}_{\ell}(k_1,k_2)
\label{eq:3D_flat_full_lens} \ea where in this case we have
defined the power spectra via $\lgl \phi (k,\vell) \phi^*
(k',\vell') \rgl = (2\pi)^2 \mathcal
{C}^{\phi\phi}(k,k',\ell)\delta_D^2(\vell-\vell')$ and $\lgl
\phi_{\ell m } (k) \phi^*_{\ell' m' } (k') \rgl =
C^{\phi\phi}_{\ell}(k,k')\delta_{\ell \ell'}\delta_{mm'}$.

One then has the following identical expressions for the flat-sky power spectra
of the shear and the convergence fields~\cite{art:Hu00}
\ba
\mathcal {C}^{\gamma \gamma}(k_1,k_2,\ell)& = &\frac{\ell^4}{4}
                                \mathcal {C}^{\phi\phi}(k_1,k_2,\ell),
\label{eq:PS_gammagamma_flat}\\
\mathcal {C}^{\kappa \kappa}(k_1,k_2,\ell) & = & \frac{\ell^4}{4}
                                \mathcal {C}^{\phi\phi}(k_1,k_2,\ell).
\label{eq:PS_conv_flat}
\ea
where now
\ba
\mathcal {C}^{\phi\phi}(k_1,k_2,\ell) = \frac{16}{\pi^2c^4}
                                \int_{0}^{\infty} k^2 \,dk
                \,I(k_1,k)I(k_2,k)
\label{eq:PS_phiphi_flat}
\ea 
with $I$ as before (see Eq.~(\ref{eq:I})).  The equations
(\ref{eq:PS_gammagamma_flat}), (\ref{eq:PS_phiphi_flat}) and
(\ref{eq:I}) open up the possibility of cosmological parameter
estimation with the shear field, whose 2-point statistics are here
related to those of the underlying gravitational potential field
(and hence to the matter density via Poisson's equation).  For
clarity, we have not considered here the practical issues of
noise, source selection function, weighting schemes, or errors in
the photometric distance indicators.  These can be dealt with, and
are studied in a companion paper~\cite{art:HeavensKitching05},
which addresses the accuracy with which cosmological parameters
can be estimated with current and future 3D weak lensing surveys.
Also considered in that paper is that one has to assume a fiducial
cosmological model in order to translate source redshifts into
comoving distance coordinates $r$.

In Figures~(\ref{fig:shearpower}) to~(\ref{fig:shearpowerderivative}) 
we illustrate the amplitude and shape of the signal produced by the 3D shear power spectrum
$\mathcal {C}^{\gamma \gamma}(k_1,k_2,\ell)$ of Eq.~(\ref{eq:PS_gammagamma_flat}) 
one may ideally expect to measure. We also show how 
the shear power spectrum varies with the value of the equation of
state parameter $w$.
The plots are for a fiducial $\Lambda$CDM cosmological model with 
parameters $\Omega_\Lambda=0.73$,
$\Omega_m=0.27$, $\Omega_b=0.27$, $H_0=71$ km/s/Mpc and equation of state of
dark energy parameter $w=-1$. Figs.~(\ref{fig:shearpower}),
(\ref{fig:shearpowerdiag}) and (\ref{fig:shearpowervsl}) show the
features of the 3D shear power spectrum by taking various cuts
through the 3D ($k_1,k_2,\ell$) space. 
We consider a survey to $r_{max}=5000$ Mpc, which replaces infinity as
the upper limit of the radial integral in Eq.~(\ref{eq:Ieq}).
The sampling interval in wavenumber $k$ is $2.5\times10^{-4}$ Mpc$^{-1}$
and in multipole ${log}_{10}(\ell)$ is $0.025$.
We also illustrate in Fig.~(\ref{fig:shearpowerderivative}) 
how the shear power spectrum varies with $w$. The differences are slight in an
individual power spectrum, but as this is a 3D study, there are many
useful $\ell$ modes, which increase the sensitivity of the 3D shear as
a cosmological parameter diagnostic.  

\begin{figure}
\includegraphics[width=0.8\columnwidth]{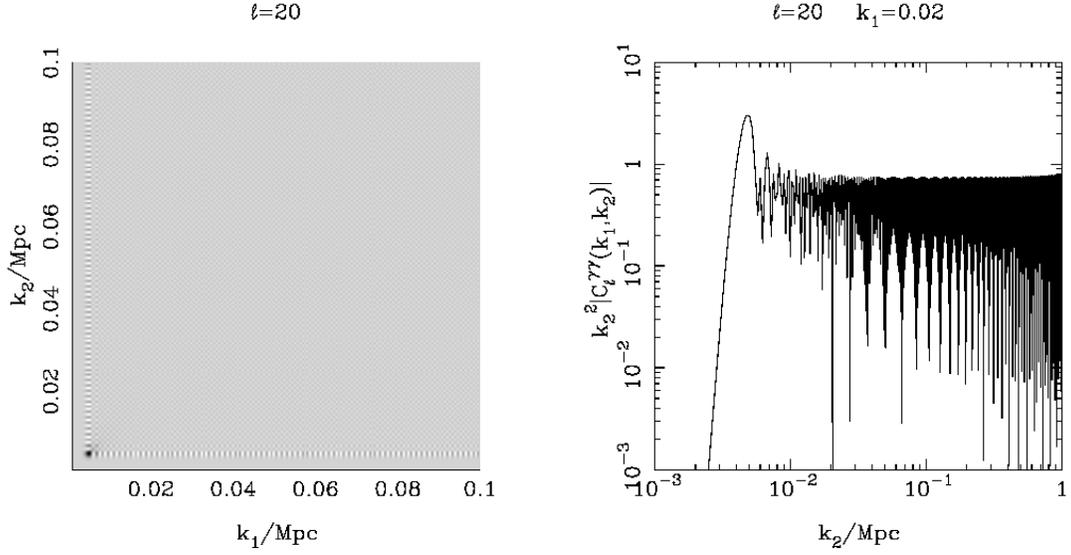}
\\
\caption{The {\it left-hand panel} shows a grayscale cross-section
through the $\mathcal {C}^{\gamma \gamma}(k_1,k_2,\ell)$ 3D 
matrix where each element has been multiplied
by $k_1k_2$. The black/white colour represents positive/negative
values. The {\it right-hand panel} represents a cut
in the $(k_1,k_2)$ plane for a given value of $\ell$ and $k_1$. 
The fiducial cosmological model parameters chosen are $\Omega_\Lambda=0.73$,
$\Omega_m=0.27$, $\Omega_b=0.27$, $H_0=71$ km/s/Mpc and $w=-1$.}
\label{fig:shearpower}
\end{figure}

\begin{figure}
\includegraphics[width=0.75\columnwidth]{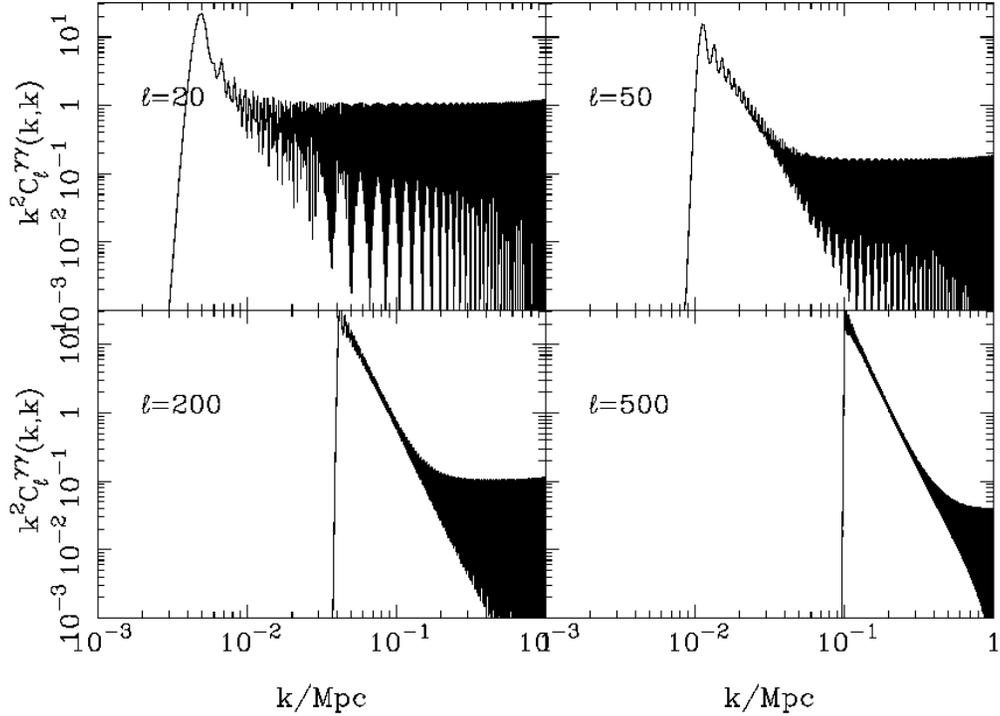}
\\
\caption{These plots show diagonal cuts in the $(k_1,k_2)$ plane at different $\ell$
values through the $\mathcal {C}^{\gamma \gamma}(k_1,k_2,\ell)$ 3D
matrix. The effect of the approximate Bessel function inequality, $kr
\geq \ell$, in Eq.~(\ref{eq:Ieq}) can be seen. 
As the $\ell$ value increases the diagonal terms of
the covariance do not become significant until $kr_{max} \approx
\ell$, where $r_{max}$ is the upper limit imposed on the $r$ integral. 
The fiducial cosmological model chosen 
is the same as in Fig.~(1).
}
\label{fig:shearpowerdiag} 
\end{figure}

\begin{figure}
\includegraphics[width=0.6\columnwidth]{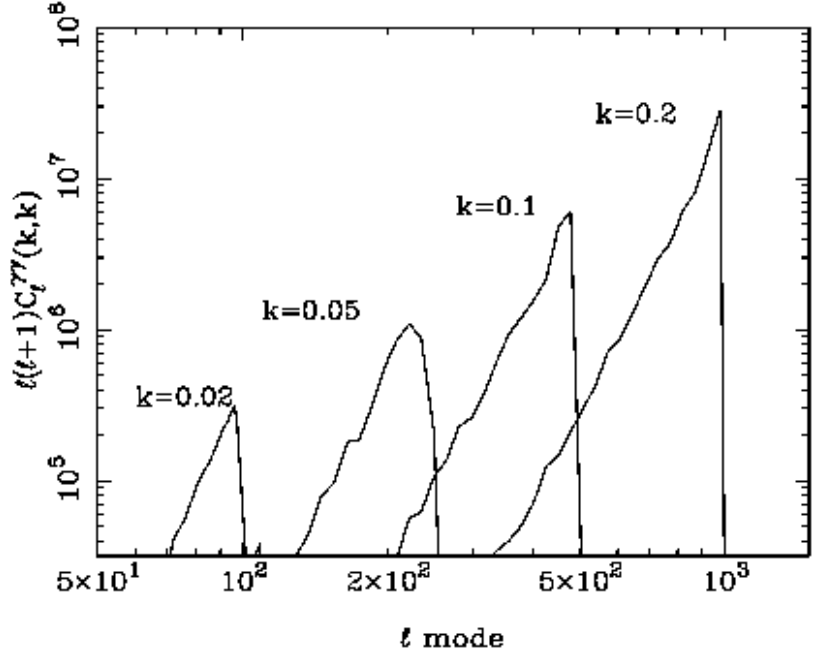}
\\
\caption{Dependence of the diagonal part
of the 3D shear power
spectrum $\mathcal {C}^{\gamma \gamma}(k_1,k_2,\ell)$ 
on $\ell$, for various $k$ values in Mpc$^{-1}$. 
The fiducial cosmological model chosen 
is the same as in Fig.~(1).}
\label{fig:shearpowervsl}
\end{figure}

\begin{figure}
\includegraphics[width=0.7\columnwidth]{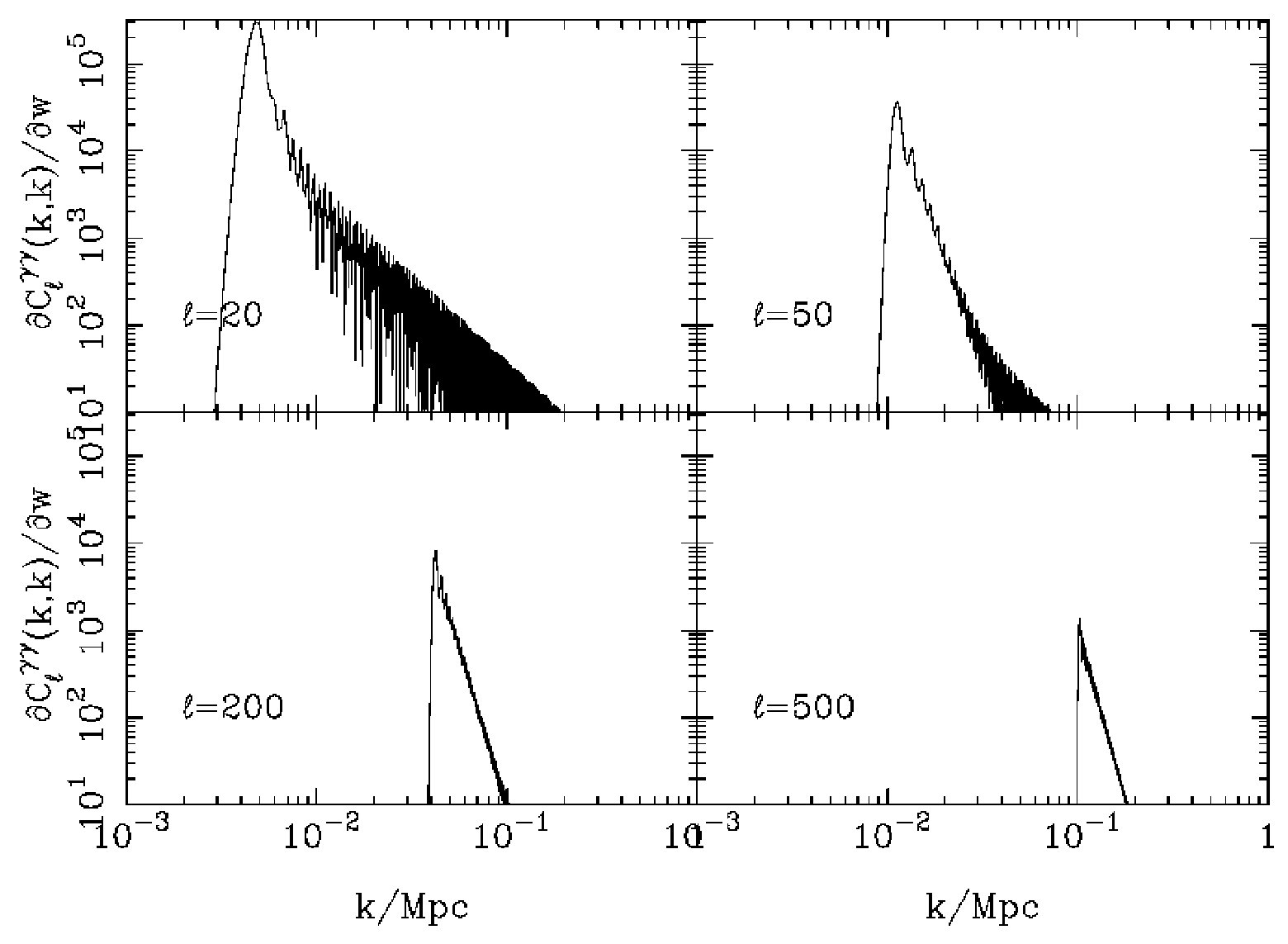}
\\
\caption{ The derivative of the diagonal elements of
the 3D covariance matrix $\mathcal {C}^{\gamma \gamma}(k_1,k_2,\ell)$ 
for various $\ell$ values, with respect to the dark energy
equation of state parameter $w$. The fiducial cosmological model chosen 
is the same as in Fig.~(1) where $w$ is allowed to vary.}
\label{fig:shearpowerderivative}
\end{figure}


\section{3D correlation function}
\label{sec:3D_corr_function}

For surveys with complicated geometry, the 3D shear correlation
function may be a more appropriate tool to use, in place of the 3D
shear power spectrum.  It is straightforward to relate the 3D shear
correlation function to the lensing potential power spectrum (and
hence the matter power spectrum), via (\ref{eq:cl_phi}), as
follows. The shear expansion as defined in Section~\ref{subsec:wl_sph_harm} implies
that
\begin{equation}
\langle\gamma(\rr)\gamma^*(\rr')\rangle = \sum_{\ell m \ell' m'}
\int dk\,dk' \langle {\,_2\gamma_{k\lm}} {\,_2\gamma^*_{k'\ell'
m'}}\rangle {\,_2Z_{k\lm}}(r,\theta,\varphi) {\,_2Z^*_{k'\ell'
m'}}(r',\theta',\varphi').
\end{equation}
Using (\ref{eq:PS_gammagamma}), this reduces to
\begin{equation}
\langle\gamma(\rr)\gamma^*(\rr')\rangle = \frac{2}{\pi}\sum_{\ell
m}\frac{1}{4}\frac{(\ell+2)!}{(\ell-2)!} \int dk\,dk' C^{\phi
\phi}_{\ell}(k,k') kk' j_\ell(kr)
j_{\ell}(k'r'){\,_2Y_{\lm}}(\theta,\varphi) {\,_2Y^*_{\ell
m}}(\theta',\varphi').
\end{equation}
Using the generalised addition theorem for spin-$s$ spherical
harmonics \cite{art:HuWhite97}, we find
\begin{equation}
\langle\gamma(\rr)\gamma^*(\rr')\rangle =
\frac{1}{2\pi}\sum_{\ell}\sqrt{2\ell+1\over
4\pi}\frac{(\ell+2)!}{(\ell-2)!} \int dk\,dk' C^{\phi
\phi}_{\ell}(k,k') kk' j_\ell(kr) j_{\ell}(k'r'){\,_2Y_{\ell,
-2}}(\beta,\alpha)e^{-2i\epsilon},
\label{eq:2_pt_corr_full}
\end{equation}
where $\beta$ is the angle between the directions $(\theta,\varphi)$
and $(\theta',\varphi')$, and $\alpha$ and $\epsilon$ are the angles
between the line joining the two points on the celestial sphere and
the lines of constant $\varphi$.  An explanatory diagram can be
found in \cite{art:HuWhite97}.  Note that one might expect that the
correlation function would depend only on the angular separation
through $\beta$, but as $\gamma$ is defined with reference to the
orthogonal coordinate system $(\theta,\varphi)$, there is an
additional dependence on the angles $\alpha$ and $\epsilon$.  Note also that
the 3D correlation function is not homogeneous.  If one
chooses, one may simplify slightly by considering pair correlations
of the shear defined with respect to axes parallel and perpendicular
to a great circle joining the two galaxies on the sky.  The rotation
of the $\gamma$ coefficients (to $\tilde\gamma$) then gives
\begin{equation}
\langle\tilde\gamma(\rr)\tilde\gamma^*(\rr')\rangle =
\frac{1}{2\pi}\sum_{\ell}\sqrt{2\ell+1\over
4\pi}\frac{(\ell+2)!}{(\ell-2)!} \int dk\,dk' C^{\phi
\phi}_{\ell}(k,k') kk' j_\ell(kr) j_{\ell}(k'r'){\,_2Y_{\ell,
-2}}(\beta,0)
\end{equation}
where $_2Y_{\ell, -2}(\theta,0)$  may be simplified to a weighted
sum of 3 associated Legendre functions.

One can derive equivalent flat-sky relations by using the
expressions~(\ref{eq:shear_flat}) for the weak-lensing shear 
on the flat-sky and the expansion~(\ref{eq:general_3D_expansion_flat_1})
for the lensing potential. Using the result
$\gamma(k,\vell) = [(\ell_x^2-\ell_y^2)/2 + i\ell_x
\ell_y]\,\phi(k,\vell)$, where $\vell$ and $k$ are defined as in the previous
section, we easily obtain
\ba
\langle\gamma(r,\vth)\gamma^*(r',\vth')\rangle =
      \frac{1}{8\pi^3}\int d^2\vell \,\,\ell^4 \int dk\, dk'
      \mathcal {C}^{\phi\phi}(k,k',\ell) kk' j_\ell(kr) j_{\ell}(k'r')\,
      e^{i \vell.(\vth - \vth')}
\ea
where $\mathcal {C}^{\phi\phi}(k,k',\ell)$ is the flat-sky power
spectra of the lensing potential. Alternatively, one can reach the
same result by expanding Eq.~(\ref{eq:2_pt_corr_full}) around the pole
of spherical coordinates $\theta\simeq0$ and taking the continuous
limit of high multipoles $\ell$.



\section{Conclusions}

In this paper we have developed in detail a complete formal study of
3D weak lensing, where one has estimates of the weak lensing shear
field at known locations in 3D space.  Most cosmological weak lensing
surveys to date have been analysed in projection on the 2D sky, where
the individual distances of the lensed galaxies are ignored. With
distance information, a more sophisticated analysis is possible, which
offers the prospect of greater statistical power. A partly 3D approach
is to divide the sources into redshift slices, a process often
referred to as tomography, but this crude division fails to explore
the complete potential provided by distance information, and it makes
sense to exploit as fully as possible the 3D information available:
one essentially has a noisy estimate of the 3D shear field at the
positions of all the source galaxies. For the estimation of some
parameters, such as the amplitude of the power spectrum, 3D
information adds relatively little, but for others, such as the
equation of state of dark energy, 3D weak lensing analysis is very
promising, and this facet of 3D lensing is explored in a companion
paper~\cite{art:HeavensKitching05}. Indeed, weak lensing on a cosmic
scale may be the best cosmological method able to answer the important
question of the nature of the dark energy.

To effect a 3D analysis, we have used the theory of spin-weight
functions successfully applied in the past to studies of the 2D CMB
polarisation on the
sky~\cite{art:ZaldSeljak97,art:Hu00,art:LewisEtAl01,art:BunnEtAl02,art:OkamotoHu03},
as well as theoretical 2D weak-lensing
analysis~\cite{art:Stebbins96,art:KamionEtAl98,art:CrittendenEtAl00}.
We have shown that the natural expansion basis to use for the 3D shear
field is the product of spin-weight 2 spherical harmonics and
spherical Bessel functions, which form a complete and orthonormal
system of tensor functions on the three-dimensional space.
Alternative expansions are clearly possible such as using tensor
spherical harmonics~\cite{art:Stebbins96} but these are more difficult
to work with and have been in some way abandoned by the CMB
polarisation community.  With our basis choice, the two-point
statistics (3D shear power spectrum, 3D correlation function) of the
weak lensing 3D shear field can be related in a straightforward manner
to the 3D power spectrum of matter density fluctuations.  The
statistics of the 3D shear field are then connected to cosmological
parameters via this relation, in addition to the redshift-distance
dependence which enters the transform. This connection is easily shown
using the {\em edth} differential operators. We are also able to
relate the power spectrum of a full-sky 3D shear expansion to the
power spectrum of a small-angle survey, where the use of spin-weight
spherical harmonics is cumbersome, and a more familiar transverse
Fourier expansion can be employed.  This is the basis for cosmological
parameter estimation using 3D weak lensing, and the formalism
presented here provides a useful theoretical framework for 3D weak
lensing studies.


\begin{acknowledgments}
PGC is supported by the PPARC through a Postdoctoral rolling
grant. TDK acknowledges a PPARC scholarship.
\end{acknowledgments}

\newpage


\appendix

\section{Spin-weight $s$ functions and spin raising and
lowering operators}
\label{appsec:spin_weight}

In this Appendix we review the notation and the mathematics
of spin-weight $s$ functions and of the geometrical spin
raising and lowering ($\edth$ and $\edthbar$)
operators defined over any two-dimensional Riemannian manifold (defined
as a 2D space with a metric).
The relation between the spin-weight functions and tensor quantities is
explained. We also clarify the relations between the $\edth$ ($\edthbar$)
operators and covariant derivatives on the 2-dimensional manifold.
We then particularise to both the 2-dimensional unit sphere and
unit Cartesian space, which will be of use to us in the main body of
this work. We finally give an overview of the spherical spin-weight spherical
harmonics. Many articles and appendices in the literature have been
dedicated to this topic (see
~\cite{art:NewmanPenrose66,art:GoldbergEtAl67,art:LewisEtAl01,art:OkamotoHu03,art:ZaldSeljak97,art:EastwoodTod82}
to cite just a useful few).

The spin-weight functions and the so-called $edth$ ($\edth$ and
$\edthbar$) operators were first introduced in the 60s by Newman
\& Penrose~\cite{art:NewmanPenrose66} and further explored by
Goldberg {\it et al.}~\cite{art:GoldbergEtAl67}. They mainly
started to be used as a convenient tool in the context of the
theory of gravitational wave radiation (see e.g. K.S.
Thorne~\cite{art:Thorne}) and later they were introduced in the
study of the Cosmic Microwave Background (CMB) all-sky
polarisation as an alternative to the tensor formalism
(see~\cite{art:ZaldSeljak97}). The similarity between the CMB
polarisation Stokes parameters $Q$ and $U$ and the weak lensing
components $\gamma_1$ and $\gamma_2$ as well as the perspective of
future large-field high resolution observations triggered recent
full-sky weak lensing studies using the spin-weight formalism
~\cite{art:Hu00,art:Hu01,art:Heavens03}.

\subsection{Representation of spin-weight functions in 2D Riemannian
manifolds}
\label{appsubsec:spin_weight_Riem}

Let us define an orthonormal basis at any point of a two dimensional
manifold with a metric $g_{ij}$
$\{ \eI ,\eII \}$.
The choice of the basis is not unique. Indeed, if we define the vectors $\m$
and $\mbar$ with respect to $\{ \eI ,\eII \}$,
the choice of the basis is described up to a phase $\psi$ such that
\ba
\m    & = & \frac{1}{{\sqrt 2}}(\eI + i \eII)
         \rightarrow e^{-i\psi} \m, \nn
\mbar & = &\frac{1}{{\sqrt 2}}(\eI -  i \eII)
         \rightarrow e^{i\psi} \mbar.
\label{eq:m} \ea The vector $\m$ and its complex conjugate $\mbar$
obey the relations $\m^j\m_j = \mbar^j\mbar_j = 0$, $\m^j\mbar_j =
1$ and $\m^i\mbar_j + \mbar^i\m_j = \delta^i_j$, where
$\delta^i_j$ is the Kronecker delta function. Here and throughout this
manuscript, we use the Einstein summation convention.

A complex function defined by $\sfunc (\nhatb) \equiv f_1(\nhatb)
+ i f_2(\nhatb)$, where $f_1$ and $f_2$ are real quantities
defined on the manifold, is said to have spin-weight $s$ if under
the transformation Eq.~(\ref{eq:m}) it transforms as $\sfunc
(\nhatb) \rightarrow e^{-is\psi} \sfunc(\nhatb)$. If we consider
an arbitrary vector on the manifold $\vb (\nhatb)$ then, for
instance, the quantities $\vb.\eI \pm i \vb.\eII$ transform as
spin-weight $\pm 1$ quantities. Generalising to a rank-s tensor
$F_{i_1...i_s}$ (where $i = 0,1$), the object
$F_{i_1...i_s}\m^{i_1}...\m^{i_s}$ transforms as a spin-weight $s$
object as each individual vector $\m^{i_n}$ contributes with a
factor $e^{-i\psi}$ when it is transformed. The rank of the tensor
is thus reflected in the transformation properties, and thus the
spin, of the corresponding complex quantity constructed.

One can therefore associate to every symmetric and trace-free
component of a rank $s$ tensor a spin-weight $s$ quantity such
that \ba \sfunc (\nhatb) & = & F_{i_1...i_s} \m^{i_1}...\m^{i_s},
\nn \smfunc (\nhatb) & = & F_{i_1...i_s}
\mbar^{i_1}...\mbar^{i_s}. \label{eq:sfunc} \ea The trace-free
condition refers to the vanishing under contraction of any two
indices in the tensor. For example, for a rank-2 tensor, it refers
to $F^i_i=F_{ii'}g^{i'i}$. As irreducible tensors of rank $s$ in 2
dimensions have only two linearly independent components, say
$F_{00...0}$ and $F_{10...0}$, we can combine them to create the
associated spin-weight $\pm s$ complex quantities $\spmfunc
(\nhatb) = F_{00...0} \pm i F_{10...0}$. It is a one-to-one
mapping. Conversely, one can express any symmetric and trace-free
rank $s$ tensor in terms of spin-weight $\pm s$ objects \be
F_{i_1...i_s} = \sfunc (\nhatb) \mbar_{i_1}...\mbar_{i_s} +
            \smfunc(\nhatb) \m_{i_1}...\m_{i_s}.
\label{eq:spin_tensor} \ee An important property of spin-weight
objects is that there exists a geometrical spin-raising (lowering)
operator $\edth$ ($\edthbar$) that have the ability of raising
(lowering) the spin-weight $s$ of an object such that under the
transformation~(\ref{eq:m}) \ba \edth   [\sfunc (\nhatb)] &
\rightarrow & e^{-i(s+1)\psi} \edth   [\sfunc (\nhatb)],\nn
\edthbar [\sfunc (\nhatb)] & \rightarrow & e^{-i(s-1)\psi}
\edthbar [\sfunc (\nhatb)]. \label{eq:edth_transf} \ea Most
relevant is that these spin-raising and lowering operators are
related to the covariant derivatives of the rank-$s$ tensor
associated to the spin-weight $s$ object on the manifold \ba \edth
[\sfunc (\nhatb)] & = &  -\sqrt{2}\m^{i_1}...\m^{i_s}\m^k \nabla_k
                                 F_{i_1...i_s}, \nn
\edthbar [\sfunc (\nhatb)]& = &   -\sqrt{2}\m^{i_1}...\m^{i_s}\mbar^k \nabla_k
                                 F_{i_1...i_s}
\label{eq:edth_sfunc_1} \ea for $s \geq 0$, and with $\m^{i_n}$
replaced by $\mbar^{i_n}$ for $s < 0$. The choice of the sign
and the normalisation are conventional but were chosen so that the form
of the spin-raising and lowering operators on the 2D sphere match
the original definition of Newman \&
Penrose~\cite{art:NewmanPenrose66} (see next section).
The covariant derivative is defined as usual.
For a rank-1 tensor $X_i$, for example, $\nabla_j X_i = \de_j X_i
- \Gamma^k_{ji}X_k$ with the Christoffel symbol depending on the
metric. If one replaces Eq.~(\ref{eq:spin_tensor}) into
Eq.~(\ref{eq:edth_sfunc_1}), one can rewrite
Eq.~(\ref{eq:edth_sfunc_1}) for $s \geq 0$ as
\ba \edth   [\sfunc
(\nhatb)] & = & -\sqrt{2} \left(\nabla_k [\sfunc (\nhatb)] \m^k +
                                 s \tau \sfunc (\nhatb) \right),\nn
\edthbar [\sfunc (\nhatb)] & = & -\sqrt{2}\left( \nabla_k [\sfunc (\nhatb)] \mbar^k -
                                 s {\overline \tau} \sfunc (\nhatb)\right)\nn
\label{eq:edth_sfunc_2}
\ea
where
\ba
\tau              & = & \nabla_k \mbar_i \m^i \m^k.
\label{eq:tau}
\ea
The equivalent relations for $s < 0$ follow from these.
One can express any differential operator in two dimensions in terms
of $\edth$ and $\edthbar$. For instance for a spin-weight s field $\,_s\,f$,
we have the following relation between the $\edth$
operators and the Laplacian operator
\ba
\nabla^2 [\,_s\,f] & = & \frac{1}{2}[\edth \edthbar + \edthbar \edth]\,_s\,f.
\label{eq:laplacian}
\ea

\subsection{Representation of spin-weight functions on the unit sphere}
\label{appsubsec:spin_weight_full}

We choose the orthonormal basis on the sphere to be aligned with
the coordinate basis vectors $\thehatb$ and $\varphihatb$ of a
spherical polar coordinate system 
$\{\etheta(\nhatb),\ephi(\nhatb) \}$ where $\nhatb$ is the radial
direction vector normal to the surface of the sphere. 
The transformation defined in Eq.~(\ref{eq:m})
corresponds to a {\it right}-handed rotation by an angle $\psi$ of this
basis around the vector $\nhatb$.
The convention chosen for the rotation is the same as in Zaldarriaga
\& Seljak~\cite{art:ZaldSeljak97} and Okamoto \& Hu~\cite{art:OkamotoHu03}
but differs from
Newman \& Penrose~\cite{art:NewmanPenrose66} and
Goldberg {\it et al.}~\cite{art:GoldbergEtAl67} so that
it conforms to the majority of CMB polarisation publications.

The metric tensor is given by \ba g_{ij} & = & \left( \begin{array}{cc}
       1  &  0\\
       0  & \sin^2 \theta
\end{array} \right)
\ea and the nonzero Christoffel symbols are \ba \Gamma_{11}^0  =
-{\sin \theta} {\cos \theta} & , & \Gamma_{01}^1  = \Gamma_{10}^1
= {\cot \theta}. \ea The tensor components of the vector $\m$ (see
Eq.~(\ref{eq:m})) are chosen to be \ba \m^k  =  \frac{1}{{\sqrt
2}}( 1,i{\csc \theta} )  & , & \m_k  =  \frac{1}{{\sqrt 2}}(
1,i{\sin \theta} ). \ea To obtain the exact relation for the
spin-raising and lowering operators $\edth$ and $\edthbar$ on the
sphere we use Eqs.~(\ref{eq:edth_sfunc_2}) and (\ref{eq:tau}). To
calculate $\tau$ we need the evaluate explicitly the covariant
derivatives $\nabla_i\mbar_j$ on the spherical basis with
coordinates $(\thehatb,\varphihatb)$. We obtain
\ba
\nabla_{\theta}\mbar_{\varphi} & = & 0, \nn
\nabla_{\varphi}\mbar_{\theta} & = & \frac{i}{{\sqrt 2}}{\cos
\theta}, \nn
\nabla_{\varphi}\mbar_{\varphi}   & = &
\frac{1}{{\sqrt 2}}{\sin \theta}\,{\cos \theta}
\ea
such that
\be
\tau = \frac{- {\cot \theta} }{{\sqrt 2}}
\ee
yielding Eqs.~(\ref{eq:edth_sfunc_sph}) and
(\ref{eq:edthbar_sfunc_sph}).
We note that the expressions obtained are identical to expressions found
elsewhere~\cite{art:NewmanPenrose66,art:ZaldSeljak97,art:LewisEtAl01,art:OkamotoHu03}.
This stems from the choice of a suitable formalism convention.
Another useful result is the full expression of the Laplacian acting
on a spin-0 quantity $\phi$
\ba
\nabla^2 \phi & = & \frac{1}{2}[\edth \edthbar + \edthbar \edth]\phi
                     = [\nablat \nablat + \csc^2 \theta \, \nablap \nablap]\phi.
\label{eq:laplacian_fullsky}
\ea
We remark that $\nabla^2$ corresponds to $-\ell(\ell+1)$ in spherical
harmonic space.

\subsection{Representation of spin-weight functions in Euclidean
space}
\label{appsubsec:spin_weight_flat}

We will also be concerned with fields defined over a 2-dimensional
Euclidean space in the so-called flat-sky approximation for which the
natural basis is the Cartesian coordinate system $\{
\ex(\nhatb),\ey(\nhatb) \}$ where $\nhatb$ is the
vector normal to the surface of the sky. In this case,
the metric tensor is given by
\ba
g_{ij} & = &
\left( \begin{array}{cc}
       1  &  0\\
       0  & 1
\end{array} \right).
\ea
The tensor components of the vector $\m$ (see Eq.~(\ref{eq:m})) are chosen to be
\be
\m^k  =  \frac{1}{{\sqrt 2}}( 1,i),
\ee
where $\m^k=\m_k$. The Christoffel symbols are all zero and thus in
Eqs.~(\ref{eq:edth_sfunc_2}) and (\ref{eq:tau}) we have that $\tau = 0$.
In this case, the differential operators $\edth$ and $\edthbar$
naturally reduce to Eqs.~(\ref{eq:edth_edthbar_flat}) \cite{art:CrittendenEtAl00}.
For completeness, the Laplacian in Cartesian coordinates reduces to
\ba
\nabla^2 \phi & = & \frac{1}{2}[\edth \edthbar + \edthbar \edth]\phi
                     = [\de_x^2 + \de_y^2]\phi.
\label{eq:laplacian_flatsky}
\ea

\section{Spin weight spherical harmonics}
\label{appsec:spin_harm}

A scalar field defined on the sphere can be expanded in spherical
harmonics $\Ylm (\theta,\varphi)$, which form a complete and
orthonormal basis on the sphere. This basis of spherical harmonics
is no longer appropriate to describe objects of spin-weight $s$.
There exists similar sets of functions defined on the sphere that
can be used instead. These are called spin-weight spherical
harmonics $\sYlm (\theta,\varphi)$  and are defined
by~\cite{art:HuWhite00} \be \sYlm (\theta,\varphi) = {\sqrt
\frac{(2\ell + 1)}{4\pi}}
                      D^{\ell}_{-s,m}(\theta,\varphi,0)
\ee where $D$ is the Wigner-D function (for technical details
see~\cite{book:VarshalovichEtAl88,art:PenroseRindler84}). Using
the previous relation one can prove that the spin-weight
$s$ spherical harmonics satisfy the relation of orthogonality 
\ba 
\int_{0}^{2\pi} d\varphi \int_{-1}^{1}d {\cos
\theta}
      \,_{s'}Y_\lmd^* (\theta,\varphi) \sYlm (\theta,\varphi)
      & = & \delta_{\ell \ell'}\delta_{m m'}\delta_{s s'},
\label{eq:sph_harm_orth_comp_1}
\ea
where $\delta_{ij}$ is a Kronecker delta function.
Also $\sYlm^*=(-1)^s\,_{-s}Y_{\ell -m}$.
A given spin-weight $s$ function on the 3D sphere can thus be 
expanded as~\cite{book:LiddleLyth00}
\ba
\sfunc (\x)      & = & \int_{0}^{\infty}d k\sum_{\ell=0}^\infty \sum_{m=-\ell}^\ell
                       [a_{s,\ell m }(k)] \sZklm(x,\theta,\varphi), \nn
a_{s,\ell m } (k)& = & \int d^3x [\sfunc (\x)]
\sZklm^*(x,\theta,\varphi). 
\label{eq:general_3D_expans} 
\ea 
where in spatially flat geometry the basis functions $\sZklm$ are given by
combinations of spherical Bessel functions and spin-weight $s$
spherical harmonics as in Eq.~(\ref{eq:basis_func_3D}).
They are orthonormal 
\be 
\int
d^3x\sZklm(x,\theta,\varphi) \,_{s'}Z_{k'\lmd}^*(x,\theta,\varphi) =
         \delta_D(k-k')\delta_{\lld}\delta_{\mmd}\delta_{s s'}
\ee
where we have used Eq.~(\ref{eq:sph_harm_orth_comp_1}) and the relation
\be
\int x^2 dx \left[ {\sqrt \frac{2}{\pi}}\, k \, j_{\ell}(kx) \right]
     \left[ {\sqrt \frac{2}{\pi}}\, k' \, j_{\ell}(k'x) \right]
     = \delta_D(k-k').
\label{eq:orth_bessel}
\ee
If we are not assuming a spatially flat space then the radial functions need to be
changed. For an open Universe for example
$j_{\ell}(x)\rightarrow X_{\ell}(\Omega_K,x)$ where $X_{\ell}$ is the
hyper-spherical Bessel function (see Liddle \&
Lyth~\cite{book:LiddleLyth00} for details).

A few properties of spin-weighted spherical harmonics when acted
by spin-lowering and rising operators (as defined in
Eqs.~(\ref{eq:edth_sfunc_sph})
and (\ref{eq:edthbar_sfunc_sph})) can be very useful
\ba 
\edth\sYlm &
= & [(\ell-s)(\ell+s+1)]^{\frac{1}{2}}
                  \,_{s+1}Y_{\lm}, \nn
\edthbar\sYlm & = & - [(\ell+s)(\ell-s+1)]^{\frac{1}{2}}
                     \,_{s-1}Y_{\lm}, \nn
\edthbar\edth\sYlm & = &  -(\ell-s)(\ell+s+1) \sYlm, \nn
\edth\edthbar\sYlm & = &  -(\ell+s)(\ell-s+1) \sYlm.
\label{eq:Prop_edth_sYlm} 
\ea 
In particular, the spin-0 and spin-2
spherical harmonics are related through 
\ba \edth \edth \Ylm
& = &\sqrt{\frac{(\ell+2)!}{(\ell-2)!}}\tYlm, \nn \edthbar \edthbar
\Ylm & = &\sqrt{\frac{(\ell+2)!}{(\ell-2)!}}\mtYlm.
\label{eq:Prop_edth_sYlm_02} \ea 
A useful consequence of these
properties is \ba \edthbar \edthbar \edth \edth \Ylm & = &
 \edth \edth \edthbar \edthbar  \Ylm =
 \frac{(\ell+2)!}{(\ell-2)!} \Ylm =
 (\ell+2)(\ell+1)\ell(\ell-1)\Ylm
\label{eq:Prop_edth_Ylm_1}
\ea
which reduces to (see e.g.~\cite{art:BunnEtAl02})
\ba
\edthbar \edthbar \edth \edth & = &
 \edth \edth \edthbar \edthbar = \nabla^2(\nabla^2 +2)
\label{eq:Prop_edth_Ylm_2}
\ea
when acting on spin-zero variables and where $\nabla$ is a covariant
derivative on the sphere.


\end{document}